\documentclass{article}

\usepackage{graphicx,latexsym}
\usepackage{amsmath,amsthm}
\usepackage{amsfonts}
\usepackage{amssymb}
\usepackage[utf8]{inputenc}
\usepackage{authblk}
\usepackage{mathtools}
\usepackage{tikz}
\usepackage{pgf,tikz,pgfplots}
\usetikzlibrary{arrows}
\usetikzlibrary{matrix}
\usepackage{caption}
\usepackage{subfigure}
\usepackage[all]{xy}

\newcommand{\R}{{\mathbb R}}  
  
\newcommand{\Z}{{\mathbb Z}}

\newtheorem{remark}{Remark}

\DeclareMathOperator{\F}{\mathcal{F}}

\pgfplotsset{compat=1.18} 

\begin{document}

\title{
Homotopy lattice gauge fields 1: \\
The fields and their properties 
}
\author[1]{
Juan Orendain
\footnote{e-mail: \ttfamily juan.orendain@itam.mx}
}
\author[2]{
Ivan Sanchez
\footnote{e-mail: \ttfamily isanchez@matmor.unam.mx}
}
\author[2]{
José A. Zapata
\footnote{e-mail: \ttfamily zapata@matmor.unam.mx}
}
\affil[1]{
Departamento de Matemáticas, 
Instituto Tecnológico Autónomo de México, 
Mexico City, Mexico
}
\affil[2]{
Centro de Ciencias Matemáticas, 
Universidad Nacional Autónoma de México, 
Morelia, Michoacan, Mexico.
}

\date{}
\maketitle

\begin{abstract} 
We introduce homotopy lattice gauge fields (HLGFs), a version of gauge fields over a discretized base, based on a notion of higher parallel transport that enriches the usual parallel transport along paths on a lattice to also consider higher dimensional paths. 

Higher dimensional data keeps information about the parallel transport along homotopies of curves. With this data, a HLGF on a base space of dimension two or three determines a principal bundle over the base manifold. This data is also responsible for our formulas for the topological charge on two-dimensional bases. 

Our framework is an application of a nonabelian algebraic topology framework developed to solve the local to global problem in higher dimensional homotopy. 

No previous knowledge of higher category theory is assumed. 

The second part will be devoted to the space of fields as an arena for doing Quantum Field Theory, and to giving the first examples of how our framework refines standard lattice gauge theory. 
\end{abstract}

\tableofcontents

\section{Introduction: The lattice homotopy cutoff, a better cutoff for gauge fields}

%

In this series of two papers, we introduce a new discretization of gauge fields. In this first part we define our gauge fields over a discrete base space, and describe their properties. The second part (work in progress) 
is devoted to the space of fields, and to showing that it is a good arena for doing Quantum Field Theory. 

Gauge fields are ubiquitous in fundamental physics. They are used to describe the so-called Standard Model of high energy physics, some systems in condensed matter physics and some approaches to quantum gravity. 

Quantum field theory requires a cutoff. Cutoffs are not considered to provide simplified frameworks; they are a necessary ingredient in the Wilsonian construction of interacting quantum field theories. 
As far as quantum physics is concerned, we could say that 
frameworks involving a cutoff are more fundamental than frameworks based on the smooth category, which is well motivated by classical considerations. 

Lattice gauge theory (LGT) is a framework for working with quantum gauge fields based on a lattice discretization of the base space. 
The usual lattice cutoff of gauge theories avoids the difficulties in producing a gauge independent procedure faced by  perturbative approaches. 
The focus is on parallel transport, and the cutoff consists of selecting $P(L) \subset P(M)$ a finitely generated subgroupoid of the path groupoid selected by an embedded lattice $L \subset M$. 
LGT is a tremendously successful framework, in particular for QCD: Its predictions about the masses of particles have been verified with accuracies higher than 99\%.%
\footnote{
More precisely, for example Dürr et al \cite{BMW:2008jgk} calculated an average proton–neutron mass to be $936$ MeV (with an uncertainty of about 3.5\%), to be compared with the experimental value of $939$ MeV. The central values differ by approximately 0.3\%.
}

It is, however, known that lattice gauge fields do not keep track of the topological information present in a $G$-gauge field in the continuum. For example, the gauge field in the continuum determines a principal $G$-bundle over the base space, and this property does not hold for lattice gauge fields (LGFs); see Section \ref{Sec2}. 
A symptom is that a formula for the topological charge that is free of ambiguities cannot be given before a continuum limit is reached; the topological charge is valued in a discrete set while the space of LGFs is a connected manifold (for a connected structure group $G$).

Now we briefly describe the contributions of this work. 
We introduce a version of gauge fields on a discretized base space endowed with a special structure designed to keep track of homotopies of curves, they are called homotopy lattice gauge fields (HLGFs). They can be seen as a particular type of ``higher gauge fields'' on a discrete base space describing a ``higher parallel transport operation''. 
We can say that the difference between HLGFs and standard lattice gauge fields is that the focus is on higher parallel transport, which includes ordinary parallel transport and keeps more information about the gauge field in the continuum associated to parallel transport along homotopies of paths. HLGFs give a (higher) algebraic structure to the previous version of homotopy aware lattice gauge fields proposed by one of the authors in collaboration with Claudio Meneses \cite{Meneses:2017vqn, Meneses:2019bok}. This algebraic structure provides the geometrical clarity that lets us uncover the parallel transport operation natural to HLGFs, and provides a coarse graining map by unambiguous gluing formulas 
(that we will introduce in the second article of this series). On a separate note, the restriction of HLGFs to dimension 2 reproduces a previous notion of higher gauge field on a discretized base proposed by Pfeiffer \cite{Pfeiffer:2003je}. 
We will see that, for base spaces of dimensions 2 or 3, a HLGF determines a $G$-bundle over the base space. We also give a simple formula to calculate the topological charge in dimension 2. 

Reading this paper does not require previous knowledge of category theory. We develop the necessary algebraic tools providing geometrical motivation. The reader will learn the essence and basic tools of a ``non abelian algebraic topology'' framework developed by Brown, Higgins and collaborators \cite{BrownNAT}. This language is also used to introduce a higher version of the gauge groupoid (or Atiyah groupoid) that gives a rich algebraic context to our higher parallel transport operation. 

Our framework can handle any smooth base manifold, and the physical motivation calls for a cutoff with a locally finite discretization (cellular decomposition or triangulation) of the base. 
In the rest of the paper, for technical simplicity, we will assume a compact base space which may have a boundary and corners and a discretization with finitely many elements. Since the framework is designed for the consistent gluing of regions, the stated technical assumption does not constitute a limitation. Another assumption is that the structure group $G$ is assumed to be connected. Our formalism works for the general case, but there are a few instances where the wording becomes unnecessarily involved for the introduction of a framework. 

An important clarification is that our main goal is to study ordinary gauge fields, not higher gauge fields as in other works that use higher category theory language. In Section \ref{GeneralHigherGaugeFields}, however, we address the relation between our work and general higher gauge fields on a lattice. 

There is a profound reason behind the ``coincidence'' of finding that our mathematical needs 
demanded by physical requirements 
had already been developed. Researchers in algebraic topology had big hopes for using higher homotopy groups to study spaces. They then realized that for any $k>1$ the groups $\pi_k X$ are abelian. Another problem they faced was the local to global problem: understanding the homotopy type of a space that is decomposed into a collection of pieces in terms of homotopy properties of its pieces, is a difficult awkward problem. Whitehead \cite{whitehead1948operators} was the first to realize that extending the study of homotopy to the study of relative homotopy leads to nonabelian structures. Later Brown, Higgins and collaborators after working for decades solved the local to global problem for relative homotopy on filtered spaces \cite{brown1979colimit}; see \cite{BrownNAT} for a pedagogical account of nonabelian algebraic topology. Filtered spaces add  structure to topological spaces (or to based topological spaces) that make it natural to replace groups with groupoids. This change was crucial for solving the local to global problem. Additionally, filtered spaces facilitate the goal of talking about relative homotopy. In physics, we are interested in local descriptions of systems, as well as in considering  relative observables, and we need a cutoff (at least as an intermediate step). In this work and in \cite{Orendain:2023tly}, we import the setting and tools of nonabelian algebraic topology to lattice gauge theory and harvest the first fruits. 

In the second part of the series, we define and give structure to the space of fields. We show that it is a good arena for QFT at a given cutoff. We provide examples, and show that our framework refines standard LGT in the sense that its predictions regarding observables living in the standard lattice are the same, but we have access to higher dimensional observables that let us keep track of homotopy data dismissed by LGT. Among the new observables that we have at our disposal is the topological charge for two-dimensional bases. We also introduce  
coarse-graining maps linking different cutoffs in such a way that an inverse limit lets us ``get rid of the cutoff''. 

The organization of this paper is as follows: In Section \ref{Sec2} we geometrically motivate the main ideas and show how the homotopy lattice cutoff brings a field in the continuum to a HLGF. Section \ref{AbstactHLGFs} introduces higher algebraic structures in a pedagogical way relying on the motivation given in the previous section. Section \ref{TopChargeSection} contains all the topological aspects of this work. We show that in dimensions 2 or 3 a HLGF determines a $G$-bundle over the base manifold, and give a formula for the topological charge in dimension two. 
The final section contains a summary and discussion.

\section{Preparing gauge fields in the continuum for the homotopy lattice cutoff}\label{Sec2}

%
%
%
%

We will describe how an ordinary gauge field in the continuum induces a higher homotopy parallel transport map. This is no more than an observation, but it is the cornerstone of the homotopy lattice cutoff and its ability to locally store information regarding the bundle structure induced by a gauge field in the continuum for base spaces of dimension 3 or 2.

 \subsection{Parallel transport}\label{PTsubsection}

We assume that the reader is familiar with the differential geometric notion of principal bundles, connections and the induced parallel transport map. We describe it briefly to set up our notation in a familiar context. Then we talk about parallel transport from an alternative point of view. 

Consider $M$ a smooth $n$-dimensional manifold with a smooth triangulation $X$ and $G$ a Lie group. 
The term curve will be used for elements of a groupoid. From now on a piecewise smooth map%
\footnote{In Section \ref{AbstactHLGFs} we introduce formally cubical and globular shapes that let us compose in higher dimensions. To relate the two shapes it is easier to parametrize curves with domain $[-1,1]$ instead of the traditional domain $[0,1]$.} 
$c: [-1,1] \to M$ will be called a singular curve on $M$%
\footnote{In the smooth category, it is more common to consider lazy paths \cite{schreiber2007parallel} as singular curves instead of piecewise smooth curves. These paths are smooth everywhere, but they are allowed to have points where 
despite of not being a constant path, in some neighborhood of the parameter space the curve is allowed to sit at a point. 
Our treatment of the smooth category is heuristic, and at this level we can describe things easier by following our conventions.}. 
Singular curves have a source and a target $s(c)= c(-1)$, $t(c)= c(1)$.

\begin{figure}[ht] \centering 
    \includegraphics{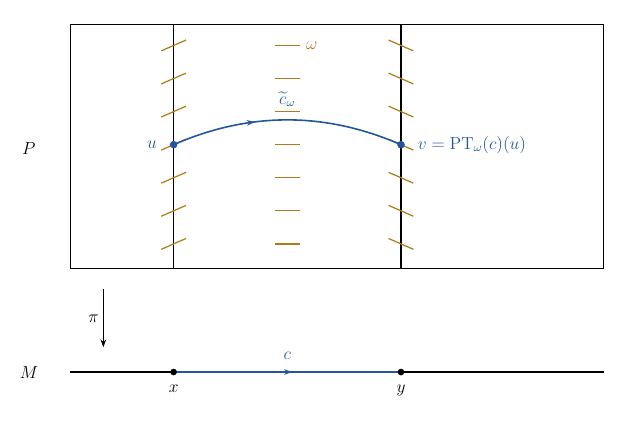}
    \caption{}
    \label{PTomega}
\end{figure}

{\em A principal bundle with a connection}. 
There may be inequivalent $G$-principal bundles $\pi: P \to M$ over $M$. 
For the moment we will choose one of them. 
A connection $\omega$ on $\pi$ determines a lift for any singular curve $c$ in $M$ once an initial condition is specified 
in the fiber over $s(c)$. 
Consider a singular curve $c$ in $M$; the lift from any initial condition induces a map $PT_\omega(c) : \pi^{-1} s(c) \to \pi^{-1} t(c)$ that commutes with the right $G$ action on $\pi$. 
We say that a connection $\omega$ determines a parallel transport map $PT_\omega$ \cite{kobayashi1996foundations}. A visual summary is given in Figure \ref{PTomega}. 

{\em Parallel transport from a connection on a bundle}. 
$\tilde{P}(M)$, the space of singular curves in $M$, has two important structures: \\
(i) It has a smooth structure that will let us talk about smooth maps with that space as domain. The idea is to consider $\tilde{P}(M)$ as a parametrized space in which a special set of parametrizations are defined to be smooth with the objective of defining a map 
$f: \tilde{P}(M) \to {\mathbb R}$ to be smooth if and only if precomposition with any of the smooth parametrizations results in a smooth map. One realization of such a smooth structure is the concept of a diffeological space \cite{iglesias2013diffeology}, which we will use when we deal with the space of fields in the second part of this series. 
 This notion of a smooth space extends the notion of a smooth manifold: All smooth manifolds are diffeological spaces. Importantly, that structure descends to quotient spaces and certain limits that we will use in the second part of this series. \\
(ii) $\tilde{P}(M)$ becomes a groupoid after taking a quotient by the equivalence relation of thin homotopy relative to vertices. It has a set of operations (source, target, composition, inverse and identities) that descend to the quotient $P(M)= \tilde{P}(M) / \sim_{thin}$ and give it the structure of a groupoid. 
The source and target maps were introduced earlier. Composition corresponds to singular curve concatenation; when it makes sense, $c_2 \circ c_1$ is another singular curve corresponding to following first $c_1$ and later $c_2$. The inverse corresponds to inverting the direction in which the singular curve is traversed $c^{-1}(t) = c(-t)$. A constant curve on $x\in M$ is called an identity and it is denoted by $id_x$. 
Thin equivalence is an equivalence relation among singular curves such that $c^{-1} \circ c \sim_{thin} id_{s(c)}$. 
For gauge fields the definition has a clear motivation: If we can go from one singular curve to another one without sweeping any area, an appropriately defined integrated curvature would vanish. 
It is a very nice equivalence relation because in some sense it is the weakest equivalence relation that makes the quotient a groupoid and because it has natural generalizations for the higher dimensional analogs of singular curves%
\footnote{In the next section, we will define thin equivalence precisely within a special context in which the concept is simply stated.}. 
The elements of $P(M)$ will be called curves or paths.

A parallel transport map $PT_\omega$ is defined on $\tilde{P}(M)$ but descends to $P(M)$. Given $c \in \tilde{P}(M)$, the induced transport map transports the fiber over the source to the fiber over the target 
$PT_\omega(c) : \pi^{-1} s(c) \to \pi^{-1} t(c)$ in a $G$-equivariant way. Additionally, given composable singular curves, $PT_\omega(c_2 \circ c_1) = PT_\omega(c_2) \circ PT_\omega(c_1)$, and $PT_\omega(c^{-1} \circ c) = id_{\pi^{-1} s(c)}$. These are the algebraic properties of parallel transport maps. 

We would like to describe $PT_\omega$ as a field with the smooth groupoid $P(M)$ as domain. $PT_\omega$ would be a groupoid homomorphism that after evaluation keeps track of the evaluation point as it happens for any field modeled as a section. 
This would provide an elegant summary of what we described above. 
The question is what is the appropriate groupoid that is the target of $PT_\omega$. 
Now we heuristically describe this idea. 
For any curve $c$, $PT_\omega(c)$'s outcome is a $G$-equivariant transport map 
$\pi^{-1} s(c) \to \pi^{-1} t(c)$, 
and we also know that the map depends on $[c]$. 
Thus, the target groupoid should have 
objects that can be interpreted as fibers over points of the base $\pi^{-1} x$ (for $x \in M$), and its 
morphisms should have the interpretation of $G$-equivariant maps $\pi^{-1} x \to \pi^{-1} y$, while remembering that they are associated to curves $c$ (with $x = s(c)$ and $y=t(c)$). 
It turns out that these clues are sufficient to define a variant
of the Atiyah groupoid, which we denote by $At(\pi)$; 
technically, it is the pullback in the category of groupoids of
the functor $q:At_{\mathrm{std}}(\pi)\longrightarrow\operatorname{Pair}(M)$ (where $At_{\mathrm{std}}(\pi)$ is the standard Atiyah groupoid)
and the endpoint functor
$E:P(M)\longrightarrow\operatorname{Pair}(M)$ (the technical terms will be introduced in Section \ref{AbstactHLGFs}).

Let us describe $At(\pi)$ with slightly more detail; a more formal introduction will be given in the following section. 
Objects are where initial conditions for parallel transport can live. There is one object for each point $x \in M$; it will be convenient to denote objects as pairs $(x, {\cal F}_x)$. 
Morphisms will dictate the parallel transport of initial conditions along paths on the base. It will be convenient to denote morphisms by pairs of the form 
$(c \in P M, T_c: {\cal F}_x \to {\cal F}_y)$ where $x$ is the source of $c$ and $y$ is the target of $c$ and $T_c$ commutes with the global right $G$ action on the fibers of $\pi$. 
The source and target maps of these morphisms are obvious. 
Identity morphisms are of the form $(id_x \in P M, id: {\cal F}_x \to {\cal F}_x)$ for all $x \in M$. 
Composition of morphisms, when defined, is also clear. 
Inversion reverses the direction of the curve and inverts the map. 

Forgetting the second entry provides a projection 
$At(\pi) \to P (M)$. 
Forgetting the first entry provides a projection to the following groupoid: Its set of objects is the set of fibers of $\pi$; that is, objects are of the form ${\cal F}_x$ for some $x \in M$. Its morphisms are maps $T_{(x,y)}: {\cal F}_x \to {\cal F}_y$ that commute with the global right $G$-action on $\pi$. The operations on this groupoid are obvious. 

In this context, a parallel transport map is a smooth groupoid homomorphism 
\[
P(M) \overset{PT_\omega}{\longrightarrow} At(\pi) , 
\]
which is a section of the mentioned projection $At(\pi) \to P(M)$.

The presentation given above starts with a connection $\omega$ on a bundle and constructs $PT_\omega$. 
A trivial, but interesting, observation is that $PT_\omega$ determines a distribution of horizontal spaces on $\pi$; that is, $PT_\omega$ determines $\omega$.

{\em Parallel transport on $(M, X_0)$}. 
The following observation is the origin of important decisions that shaped the work presented in this article. It suggests working on a more economical space of singular curves in $M$. Parallel transport on the restricted space of paths still determines the connection $\omega$ (up to certain gauge transformations), while leading to simpler gauge transformations and a conceptually simpler framework. 

Let $\tilde{P}(M, X_0) \subset \tilde{P}(M)$ be 
the space of piecewise smooth curves $c: [-1,1] \to M$ with source and target points, $s(c) = c(-1)$, $t(c) = c(1)$, restricted to lie in $X_0$ the set of vertices of the triangulation $X$. 
$P(M, X_0)= \tilde{P}(M, X_0) / \sim_{thin}$ is a subgroupoid of $P(M)$. See Figure \ref{P(M,X0)}. 

\begin{figure}[ht] \centering
    \includegraphics{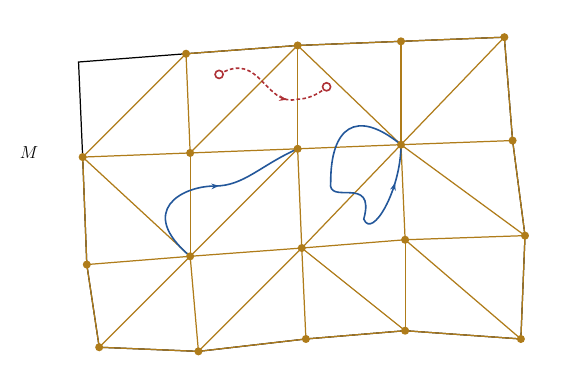}
    \caption{The blue singular curves belong to $\tilde{P}(M, X_0)$, and the red does not.}
    \label{P(M,X0)}
\end{figure}

The restricted parallel transport map $PT_\omega|_{\tilde{P}(M, X_0)}$ also descends to the quotient $P(M, X_0)$. This groupoid has a discrete set of objects $X_0 \subset X$. On the other hand, the set of paths (morphisms) has not acquired extra restrictions apart from having restricted endpoints. 

A question arises: Does $PT_\omega|_{\tilde{P}(M, X_0)}$ determine $\omega$? \\
Geometric intuition tells us that it should be enough to distinguish between any two different connections (up to gauge transformations that are the identity over points in $X_0$): 
If the two connections have different curvature, there must be a point on the base space and a neighborhood of that point where parallel transport along small loops would detect the difference in curvature. The mentioned small loop can be extended to have a base point in $X_0$; this path would be able to distinguish (modulo gauge) the local features of the two connections under consideration. Since 
the two considered connections must also have the same 
parallel transport along noncontractible loops, they must completely agree. 

Before proceeding or giving any details notice that since $X_0$ is a discrete set, information about the fibers over $X_0$ cannot encode the structure of a bundle over $M$. Thus, parallel transport along paths in $P(M, X_0)$ is an operation that does not rely on a preexisting bundle structure. Then parallel transport on $P(M, X_0)$ lets us distinguish between any two connections on a given bundle and, at the same time, permits a framework in which connections on any possible $G$ bundle over $M$ are treated at once. It is natural to conjecture that $PT_\omega|_{\tilde{P}(M, X_0)}$ can also distinguish between different bundles. 

The heuristic argument of the previous paragraph was transformed into a rigorous construction by Barrett \cite{Barrett:1991aj} following a comment from Kobayashi \cite{kobayashi1996foundations}. An explicit construction of the bundle and the connection from the parallel transport map is in \cite{Barrett:1991aj} 
(considering a base with a single selected base point $(M, \ast)$). A ``groupoid version'' that is more directly related to our exposition can be found in \cite{Meneses:2019bok}%
\footnote{The construction given in \cite{Meneses:2019bok} uses a triangulation arising from barycentric subdivision, but it is easily adapted to 
our situation where we have $X$ the triangulation of a manifold 
with compatible vertex orders, regarded as a simplicial set.}. 

In summary, the Barrett-Kobayashi construction tells us that the map 
\[
(\pi , \omega) \leftrightarrow PT|_{\tilde{P}(M, X_0)} 
\]
is invertible (up to gauge).

\subsection{Higher homotopy parallel transport in the continuum}\label{HHPTcontinuum}


Here we heuristically describe how an ordinary gauge field acts on homotopies of curves on a base space. We call the resulting notion ``higher homotopy parallel transport''. A single curve lets the gauge field parallel transport initial conditions on the fiber over its source point, to final conditions on the fiber over its target point. Homotopies of curves let the gauge field transport homotopies of initial conditions, and yield homotopies of final conditions. If we assign globular shapes to the homotopies of initial conditions and of curves and of final conditions, the resulting higher homotopy parallel transport operation has powerful gluing and composition properties. The transport of globular homotopies of initial conditions of dimension 1 captures non trivial information about the gauge field. 
In Section \ref{TopChargeSection} we will see that there is no interesting information to capture from the transport of 2 dimensional homotopies of initial conditions, and that the transport of k dimensional homotopies of initial conditions may contain interesting information about the gauge field, but the strict globular machinery that we use does not capture it for $k \geq 3$.

\begin{figure}[ht] \centering 
    \includegraphics{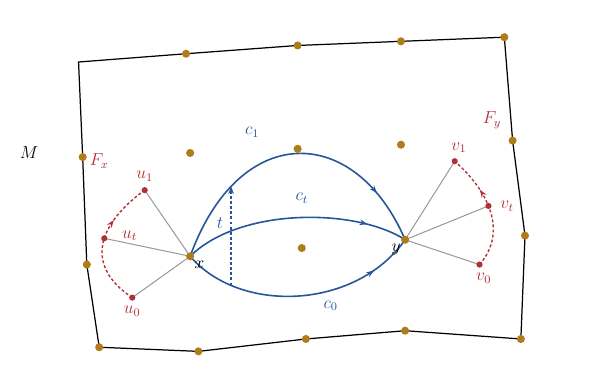}
    \caption{A homotopy of singular curves is a singular 2-globe sharing endpoints. The gauge field transports homotopies of initial conditions over the source along a singular 2-globe to a homotopy of final conditions over the target.}
    \label{PT2-globos}
\end{figure}

Consider $PT$ a smooth parallel transport map on $(M, X_0)$. 
A homotopy of curves $\{ c_t \in \tilde{P}(M, X_0) \}_{[0,1]}$ sharing their source and target points in $X_0$ will be called a singular 2-globe. 
Now consider an initial condition $u\in {\cal F}_{s(c_t)}$ in the fiber over $s(c_t) \in X_0$. The parallel transport along the singular 2-globe 
$\{ c_t \}_{[0,1]}$ leads to $\{ PT(c_t)[u] \}_{[0,1]}$ a homotopy in ${\cal F}_{t(c_t)}$ interpolating between 
$PT(c_0)[u] \in {\cal F}_{t(c_t)}$ and $PT(c_1)[u] \in {\cal F}_{t(c_t)}$. 
Notice that this transport operation started with an element of one type (a single initial condition) and ended up with an element of another type. 
On the other hand, a homotopy of points in the initial fiber $\{ u_t \in {\cal F}_{s(c_t)} \}_{[0,1]}$ is transported along the singular 2-globe to a homotopy of points in the final fiber 
\[
PT (\{ c_t \}_{[0,1]}) [ \{ u_t \}_{[0,1]}] = 
\{ PT(c_t)[u_t] \in {\cal F}_{t(c_t)} \}_{[0,1]} . 
\]
In summary, $PT(\{ c_t\}_{[0,1]}) 
\doteq \{ PT(c_t)\}_{[0,1]}$ is a homotopy interpolating between $PT(c_0)$ and $PT(c_1)$ 
taking homotopies of initial conditions on ${\cal F}_{s(c_t)}$ and transporting them to homotopies of final conditions on ${\cal F}_{t(c_t)}$. Our objective is to interpret $PT(\{ c_t\}_{[0,1]})$ as a transport map in itself acting on the appropriate type of initial conditions. See Figure \ref{PT2-globos}. 

Now that we know what kind of objects can be transported by a singular 2-globe to objects of the same type, let us see if the possible compositions of singular 2-globes get along with parallel transport. 

Singular 2-globes can be composed in two directions. See Figure \ref{+i} in the next section for a self explanatory picture. 
Notice that singular 2-globes have 0-dimensional source and target points, and they also have 1-dimensional source and target singular curves. 
First, singular 2-globes 
$\{ c^1_t \}_{[0,1]}, \{ c^2_t \}_{[0,1]}$ such that the 0-dimensional target of $c^1$ coincides with the 0-dimensional source of $c^2$ can be composed 
$\{ (c^2 \circ_0 c^1)_t \}_{[0,1]}= \{ c^2_t \circ c^1_t \}$. 
Parallel transport along the resulting 2-globe is, clearly, the composition of the parallel transport operations along the original 2-globes. 
Second, composition of $c^1$ and $c^2$ is possible when the 1-dimensional target of $c^1$ coincides with the 1-dimensional source of $c^2$ (i.e. if $c^1_1 = c^2_0$); we write 
$\{ (c^2 \circ_1 c^1)_t \}_{[0,1]}$ 
for the concatenation of homotopies relative to fixed endpoints by lifting the condition of being fixed of the middle point (and redefining the parameter $t$). 
Parallel transport along $\{ (c^2 \circ_1 c^1)_t \}_{[0,1]}$ acting on homotopies in the initial fiber of the type $\{ (u^2 \circ u^1)_t \}_{[0,1]}$ (where $(u^2 \circ u^1)_t$ denotes the class of a curve in ${\cal F}_{s(c_t)}$ resulting from composing representatives of $u^1_t , u^2_t$) is a distributive operation as stated below. 
In summary, these two composition operations of singular 2-globes are compatible with parallel transport. Since the parallel transport map $PT$ in the continuum is assumed to be smooth, in particular it is continuous leading to compatibility as stated by the simple algebraic rules 
\[
PT (\{ (c^2 \circ_0 c^1)_t \}_{[0,1]}) = 
PT (\{ c^2_t \}_{[0,1]}) \circ PT (\{ c^1_t \}_{[0,1]}) , 
\]
\[
PT (\{ (c^2 \circ_1 c^1)_t \}_{[0,1]}) [ \{ (u^2 \circ u^1)_t \}_{[0,1]} ] = 
\{ (v^2 \circ v^1)_t \}_{[0,1]} , \mbox{ with } v^i_t = PT(c^i_t)[u^i_t] . 
\]

Analogously, $PT$ acts on higher homotopies of curves leading to maps parallel transporting higher homotopies of initial conditions. To have convenient algebraic gluing rules, we must choose a shape; we will continue to use globes.%
\footnote{A cubical structure is another choice; the resulting formalism is transparent and powerful \cite{Orendain:2023tly}. The globular choice is more economical, and the notion of gauge transformations is simpler.}
Higher homotopies of curves in the base will be higher dimensional globes, and initial conditions on the fiber over the source point will be globes of the corresponding dimension.

To sharpen the presented ideas, we introduce a higher homotopy extension of the Atiyah groupoid. Here we will not be rigorous yet; what we present here is a prelude to what we will introduce formally in the following section. 

We want a higher dimensional extension of $P(M, X_0)$ describing homotopies of paths in $P(M, X_0)$. In the globular structure: 
The objects would be points in $X_0$, the 1-morphisms would be paths in $P(M, X_0)$, the 2-morphisms would be 2-globes interpolating between 1-morphisms, etc. After taking a quotient by thin homotopy, singular k-globes become elements of a k-groupoid, and are called k-globes. The collection of k-globes for all $k\geq 0$ get organized into the structure of a strict infinity groupoid \cite{brown2007new}. 
A singular (k+1)-globe can be regarded as a homotopy of k-globes sharing their source and target (k-1)-globes; in the next section, we give a formal definition in a restricted context. 
Clearly, the task of making this construction rigorous in the smooth category is nontrivial; see \cite{schreiber2007parallel}. We will define the appropriate notion of thin homotopy in our context in the next section. 
For our heuristic description, let us use the notation $\rho^\bigcirc(M, X_0)$ and $\rho^\bigcirc {\cal F}_x$ for the infinity groupoid of globes on $M$ (to be constructed in the next section) and for the fundamental homotopy infinity groupoid of the fiber ${\cal F}_x$ \cite{brown2007new} respectively. 
We would like to enrich $\rho^\bigcirc(M, X_0)$ for describing the higher homotopy parallel transport described above, where the initial conditions that may be transported are higher homotopies in the fibers over $X_0$. Thus, there should be one object per vertex $x \in X_0$, but they should be thought of as $\rho^\bigcirc {\cal F}_x$, and the morphisms should have the interpretation of describing globular higher homotopies of parallel transport maps, which means commuting with the right $\rho^\bigcirc G$ action on the set of objects $\{ \rho^\bigcirc {\cal F}_x \}_{x \in X_0}$. Let us call the resulting structure $At^\bigcirc(M, X_0; \rho^\bigcirc G)$. 

In this language, we have heuristically described that a smooth parallel transport map $PT$ in the continuum induces a higher homotopy parallel transport map 
\[
\rho^\bigcirc(M, X_0) \overset{PT}{\longrightarrow} At^\bigcirc(M, X_0; \rho^\bigcirc G) , 
\]
which is a section of the projection $At^\bigcirc(M, X_0; \rho^\bigcirc G) \to \rho^\bigcirc(M, X_0)$. This picture will be described with some detail in the following section.

Physical motivations call for a cutoff in quantum field theories. The homotopy lattice cutoff sketched below leads to a context of parallel transport along restricted path homotopies. That reduced context will let us give precise and simple definitions of the notions heuristically described above.

\subsection{The homotopy lattice cutoff: Essential ingredients}
\label{Key ingredients}

{\em The cutoff}. 
Cutoffs are not considered to provide simplified frameworks; they are a necessary ingredient in the Wilsonian construction of interacting quantum field theories. 
We could say that as far as quantum physics is concerned, cutoff frameworks for field theory are more fundamental than a formalism in the smooth category, which is well motivated by classical considerations. 
The usual lattice cutoff of gauge theories avoids the difficulties in producing a gauge independent procedure faced by  perturbative approaches. 
The focus is on parallel transport, and the cutoff consists of selecting $P(L) \subset P(M)$ a finitely generated subgroupoid of the path groupoid selected by an embedded lattice $L \subset M$. This  determines a space of gauge fields in the lattice and a cutoff map 
\[
c_L: 
\mbox{Gauge fields in the continuum} \to \mbox{Gauge fields in the lattice} . 
\]

Our goal in this section is to describe the homotopy lattice cutoff of gauge theories, which refines the standard lattice cutoff as we explain later on. 
The focus is on higher parallel transport along path homotopies selected by a cellular decomposition $X$ of $M$. This determines a space of gauge fields in the homotopy lattice and a cutoff map
\[
c_X: 
\mbox{Gauge fields in the continuum} \to \mbox{Gauge fields in the homotopy lattice} . 
\]

\medskip

\begin{figure}[ht] \centering
    \includegraphics{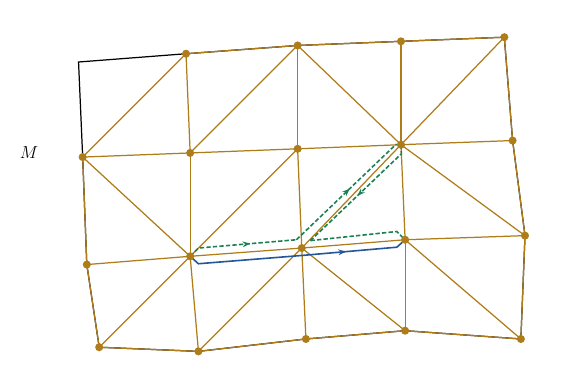}
    \caption{}
    \label{ThinEq}
\end{figure}

{\em The fundamental higher homotopy groupoid of a filtered space}. 
It is an appropriate finitely generated subinfinity-groupoid of $\rho^\bigcirc(M, X_0)$ that will let us extend the cutoff $P(L)=P(X_1) \subset P(M)$ leading to standard lattice gauge fields. A truncation should lead to a rigorous, powerful and simple formalism. It may look like a lot to ask, but it turns out that we are looking for a structure that already exists: It is $\rho^\bigcirc(M, X_\ast)$ the infinity fundamental groupoid of the filtered space $(M, X_\ast)$. 
A filtered space is a topological space together with a collection of nested subsets such that their union equals the space $X_0 \subseteq X_1 \subseteq \ldots \subseteq X_n = M$. We are interested in skeletal filtrations; that is, the space $M$ is assumed to have cellular decomposition $X$, and the subset $X_k$ of the filtration is the k-th skeleton of $X$. 
A cubical version was developed  by Brown and Higgins starting in the early 80s \cite{Algebraofcubes} (see also the book \cite{BrownNAT}), and \cite{brown2007new} gives a globular version that will be essential in this paper. We will describe a minimal version of it in the next section. 
The truncation consists of not allowing singular globes to roam everywhere in $M$: Singular k-globes  are restricted to have image in $X_k$ the k-skeleton. 
The enormous simplification, as compared with the construction of $\rho^\bigcirc(M, X_0)$ in the smooth category, comes from using the skeletal structure to define thin equivalence: Two singular k-globes are thin equivalent if there is a homotopy interpolating between them with the property that the image of any of the j-cells of the k-globe is contained in $X_j$ for any $j\leq k$. Figure \ref{ThinEq} shows two singular curves sharing endpoints that are thin equivalent. 

{\em The higher homotopy Atiyah groupoid of a filtered space}. 
Recall the heuristic description of $At(\pi)$ given in Section \ref{PTsubsection}. 
Now we present a variant of $At(\pi)$ that provides a framework for describing our higher homotopy parallel transport with internal group $G$. Let us call it $At^\bigcirc(X_\ast, \rho^\bigcirc G)$. 
There are three important changes: 
The first is that we reduced the set of allowed base points to be $X_0 \subset M$. As discussed earlier, this implies that the set of fibers over the base points does not know the bundle structure. 
The second is the cutoff induced by $X_\ast$, which carries a simple definition of thin equivalence. 
The third is that the goal of the structure is to parallel transport homotopies of elements of the fibers ${\cal F}_x$ along appropriate homotopies of paths. 
The resulting structure has objects and morphisms denoted by pairs. The second entry of the objects $(x, \rho^\bigcirc_k {\cal F}_x)$, where the initial conditions to be parallel transported live, is appropriate for describing $k$-dimensional globular homotopies in ${\cal F}_x$ (for some $k$). 
The first entry of the morphisms $(c \in \rho^\bigcirc_{k+1} X_\ast, T_{c}: \rho^\bigcirc_k {\cal F}_x \to \rho^\bigcirc_k {\cal F}_y)$ is a $k+1$ globe with 0-dimensional source $x \in X_0$ and 0-dimensional target $y \in X_0$; 
the second entry is a map commuting with the right $\rho^\bigcirc_k G$ action. The first entry of those morphisms models a homotopy of curves, and the second one models the corresponding homotopy of transport maps appropriate for acting on homotopies of initial conditions. 

\begin{figure}[ht] \centering
    \includegraphics{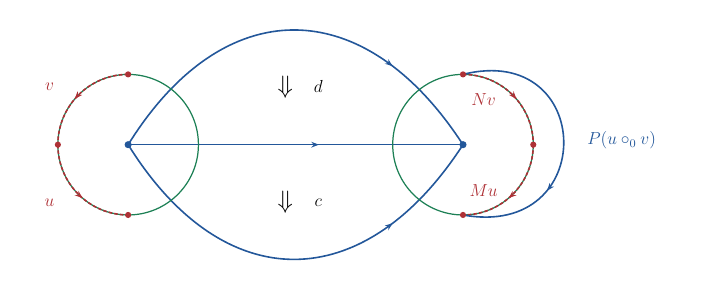}
    \caption{}
    \label{CompatAt}
\end{figure}

Importantly, the homotopy structures of the objects and that of the morphisms are coordinated precisely as needed for us to describe homotopy parallel transport. The conditions are stated at a technical level, but the idea is simple. Now we describe an instance of these conditions through an example depicted in the Figure \ref{CompatAt}: Consider a situation in which we have $G=U(1)$ and initial conditions at $x$ that are a homotopy (relative to fixed endpoints) written as the composition $u \circ_0 v \in \rho^\bigcirc_1 S^1_x$. The initial condition is transported along a 2-globe that is also expressed as a composition $c \circ_1 d \in \rho^\bigcirc_2 X_\ast$. 
The morphisms of the groupoid that we are constructing are associated with pairs; in the described situation we have three different pairs $(c , M)$, $(d , N)$ and $(c \circ_1 d , P)$. The compatibility condition says that we can transport the glued initial condition $u \circ_0 v$ using $P$, or we can transport the unglued pieces $u, v$ using $M, N$ respectively and then glue them, obtaining the same result 
\[
P (u \circ_0 v) = (M u) \circ_0 (N v) . 
\]
Similar conditions ensure that if we have initial conditions at $x$ corresponding to a homotopy $u \in \rho^\bigcirc_1 S^1_x$ that are transported along a 2-globe $c \in \rho^\bigcirc_2 X_\ast$, then we can calculate the initial point (or source) of the transported homotopy in two equivalent ways: We can calculate the source of $u$ and transport it using the 1-dimensional source of $c$, or we can first transport $u$ using $c$ and then calculate its source. Etc.

\subsection{The homotopy lattice cutoff: A preliminary example}
\label{PrelimExample}

\begin{figure}[ht] \centering
    \includegraphics{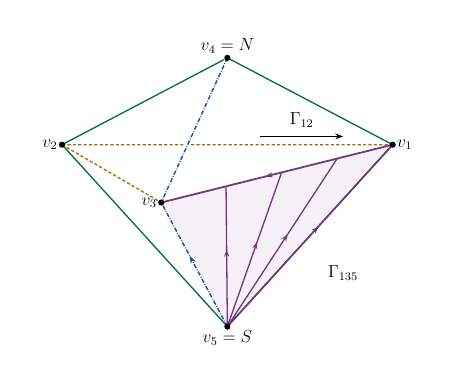}
    \caption{The triangulation, the 1-globe $\Gamma_{12}$ and the 2-globe $\Gamma_{135}$.}
    \label{TriangS2+globes}
\end{figure}

As a preliminary example, consider the case where $M = S^2$ and $G=SO(2)$. See Figure \ref{TriangS2+globes}. 
The truncation is induced by a triangulation $X$. The triangulation has a set of vertices denoted by $X_0$, a 1-dimensional skeleton consisting of edges and vertices and denoted by $X_1$ and a two-dimensional skeleton consisting of faces, edges and vertices  denoted by $X_2$. The skeletal filtration induced by the triangulation is $X_\ast = ( X_0 \subseteq X_1 \subseteq X_2 = M)$. As shown in Figure \ref{TriangS2+globes}, our triangulation $X$ has five vertices; two of them located on the North and South Poles and other three vertices on the Equator located at equidistant points. The Greenwich meridian lies on $X_1$; it is the curve joining the North and the South poles visiting vertex $v_3$, which is the intersection with the Equator (a triangle with vertices $v_1, v_2, v_3$). 
As mentioned earlier, we need a triangulation associated with an underlying simplicial set. To provide this structure we number the vertices: The only vertices that have not been assigned a number are $v_4 = v_N$ and $v_5 = v_S$ corresponding to the North and South poles, respectively. This completes the description of $X_0$. 
The set of edges is 
$\{ [v_1, v_2], [v_2, v_3], [v_1, v_3], [v_1, v_4], [v_2, v_4], [v_3, v_4]$, 
$[v_1, v_5], [v_2, v_5], [v_3, v_5] \}$. 
Notice that we have written 1-simplices, edges, as a pair of vertices respecting the order of vertices. 

A set of generators of the path groupoid in $X_1$ is $\{ \Gamma_{ij} \}$ with $i<j$, where $\Gamma_{ij}$ represents a singular curve with image $[v_i, v_j]$ and running in the opposite direction. That is, $s(\Gamma_{ij}) = v_j$ and $t(\Gamma_{ij}) = v_i$.  This convention is compatible with the conventions followed in \cite{Meneses:2019bok}, as explained in Section \ref{AbstactHLGFs}. 
If we want to refer to a singular curve running in the opposite direction we write $-\Gamma_{ij}$. 
The set of faces is $\{ [v_1, v_2, v_4], [v_2, v_3, v_4], [v_1, v_3, v_4], [v_1, v_2, v_5], [v_2, v_3, v_5], [v_1, v_3, v_5] \}$ (where triplets of vertices are written respecting their order). 
A set of generators of the 2-groupoid of 2-globes in $X_2$ is $\{ \Gamma_{ijk} \}$ with $i<j<k$, where $\Gamma_{ijk}$ represents a singular 2-globe going from the singular curve $\Gamma_{jk}$ to the singular curve $(-\Gamma_{ij}) \circ \Gamma_{ik}$. The specific homotopy of singular curves $c_t$ is shown in Figure \ref{Gamma{ijk}As2globes} (in Section \ref{Globes+ThinHomot}). Additionally, a concrete formula is given in \cite{Orendain:2023tly}; again, we adopt this convention to be compatible with \cite{Meneses:2019bok}.

If we want to refer to a singular 2-globe with source and target singular curves running in the opposite direction, or what is the same with 0-dimensional source and target interchanged, we write $-_0\Gamma_{ijk}$, and $-_1\Gamma_{ijk}$ denotes the singular 2-globe where 1-dimensional source and target  have exchanged places. 
This completes the description of the triangulation $X$, the induced skeletal filtration $X_\ast$ and our conventions for associating generating singular globes to simplices.

The infinity groupoid $\rho^\bigcirc(S^2, X_\ast)$ has three nontrivial levels $\rho^\bigcirc_0(S^2, X_\ast) = X_0$ the objects (or vertices), $\rho^\bigcirc_1(S^2, X_\ast) = P(X_1)$ the path groupoid and $\rho^\bigcirc_2(S^2, X_\ast)$ the 2-groupoid of 2-globes.

Now we will undertake the task of describing one given parallel transport map using the example of $PT_{\rm round}$ the parallel transport of unit vectors over points in $X_0$ resulting from the Levi-Civita connection associated with the metric of round unit sphere $S^2$. 

The situation is geometrically clear, but to describe the action of $PT_{\rm round}$ we need the reference provided by  a trivialization on the fibers over $X_0$. Parallel transport along a curve will correspond to an element of $SO(2)$, and parallel transport along a singular 2-globe will correspond to a homotopy of parallel transport maps (considered up to relative homotopy) described by a curve in $SO(2)$ considered up to homotopy relative to having fixed endpoints. 
We will choose an arbitrary trivialization over the north pole $v_4 = v_N$. In order to help us give a simple description, the trivialization on the fibers over the other vertices will be chosen using $PT_{\rm round}$. The trivializations over $v_1 , v_2$ and $v_3$ are chosen transporting the trivialization from $v_4$ using $PT_{\rm round}$ along the link that joins them. The local trivialization over $v_5 = v_S$ is chosen transporting the trivialization on $v_4 = v_N$ with $PT_{\rm round}$ along the Greenwich meridian.

\begin{figure}
    \centering
    \includegraphics{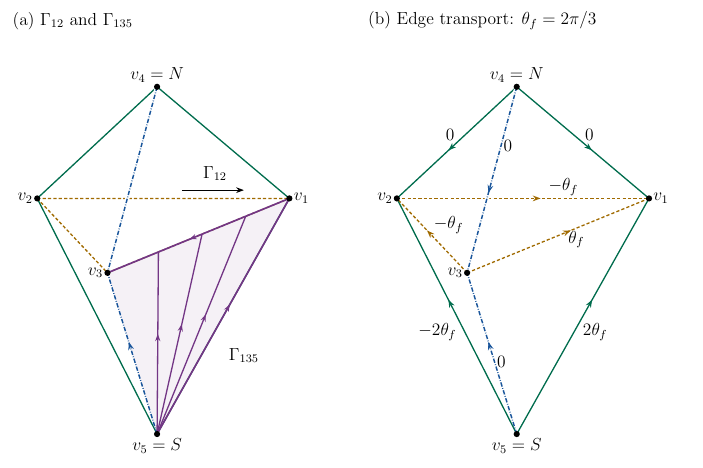}
    \caption{(a) The triangulation, the 1-globe $\Gamma_{12}$ and the 2-globe $\Gamma_{135}$. (b) Parallel transport on the round sphere.}
    \label{PTs2}
\end{figure}

According to the convention described above, parallel transport on paths corresponding to edges is 
\begin{equation}\label{PTedges}
	PT_{\rm round}(\Gamma_{ij}) = R(n_{ij} \theta_f) \in SO(2) \ \mbox{ with } \theta_f =\frac{4\pi}{6} , 
\end{equation}
for certain integer numbers $n_{ij}\in {\mathbb Z}$, as shown in Figure \ref{PTs2}. When the arrow in the figure is opposite to $\Gamma_{ij}$, we invert the sign of the coefficient of $\theta_f$ shown in the figure to compute $n_{ij}$. 

Homotopy parallel transport along singular 2-globes, as mentioned above, is described by curves in $SO(2)$ considered up to homotopy relative to having fixed endpoints. In our notation, such classes of curves in $SO(2)$ are denoted by a triple $(R \in SO(2), R' \in SO(2); x \in [0,1])$ satisfying $R' =
\exp(i x) R$. The triple determines the singular curve $c(t)= \exp(i t x) R$, and $(R, R'; x)$ denotes the homotopy class of the curve relative to having fixed endpoints. 

In our example, homotopy parallel transport along singular 2-globes associated with faces is 
\begin{equation}\label{PTfaces}
	PT_{\rm round}(\Gamma_{ijk}) = (R(n_{jk} \theta_f) , R((-n_{ij}+n_{ik}) \theta_f) ; x_{ijk}) , 
\end{equation}
with $x_{124}= \theta_f$, $x_{234}= \theta_f$, $x_{134}= -\theta_f$, $x_{125}= -\theta_f$, $x_{235}= -\theta_f$, $x_{135}= \theta_f$.

\begin{figure}[ht] \centering
    \includegraphics{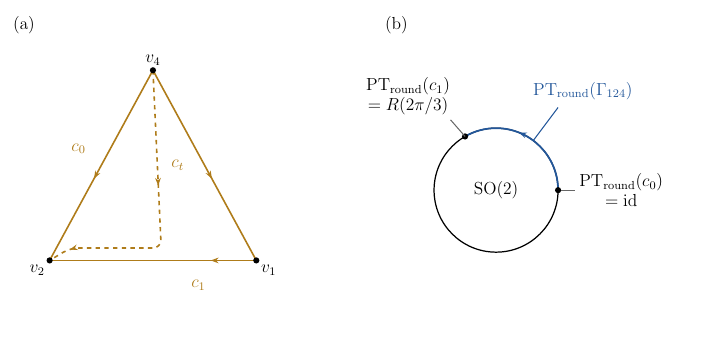}
    \caption{The first panel shows a 2-globe as a homotopy of curves sharing endpoints. In the second panel, the transport map acts on a 2-globe, shown as interpolating between 1-globes. The outcome is shown as the blue curve up to homotopy rel. to endpoints.}
    \label{HomotopyPT2globos}
\end{figure}

A visual representation of homotopy parallel transport on 2-globes is shown in Figure \ref{HomotopyPT2globos}. 

Parallel transport along any singular curve contained in $X_1$ is calculated by composing parallel transport along the edges. Similarly, as we will show in Section \ref{AbstactHLGFs}, homotopy parallel transport along any singular 2-globe fitting in $X_2$ can be calculated composing homotopy parallel transport along singular 2-globes associated with faces and along degenerate singular 2-globes with image contained on the edges. 
As an example, consider singular 2-globes interpolating between the Greenwich meridian and meridians at longitude $\frac{2\pi}{3}$, $2\frac{2\pi}{3}$ and $2\pi$; see Figure \ref{PTs2}. 
Let us call the described singular 2-globes $c_\lambda$ with $\lambda \in \{ \frac{2\pi}{3}, 2\frac{2\pi}{3}, 2\pi \}$. Because the metric on the unit round sphere has constant curvature, we know that 
\begin{eqnarray*}
	PT_{\rm round}(c_{\frac{2\pi}{3}}) &= &(id , R(2\theta_f) ; 2\theta_f) , \\
	PT_{\rm round}(c_{2\frac{2\pi}{3}}) &= &(id , R(4\theta_f)= R(\theta_f) ; 4\theta_f) , \\
	PT_{\rm round}(c_{3\frac{2\pi}{3}}) &= &(id , R(6 \theta_f)= id ; 6\theta_f = 4\pi) . 
\end{eqnarray*}

Notice that {\em for any parallel transport map} $PT$ the last entry of the above evaluations must have the structure \\
$PT(c_{2\pi}) = (PT(\mbox{Greenwich}) \in SO(2) , PT(\mbox{Greenwich}) ; x_{c_{2\pi}})$, and this requires that 
\[
Q(PT) = \frac{1}{2\pi}
x_{c_{2\pi}} 
\]
is an integer. 
This integer is called the topological charge of $PT$; in the case of the connection induced by the round metric on $S^2$, we obtain $Q(PT_{\rm round}) = 2$. 
In Section \ref{TopChargeSection}, we give more details about the calculation of the topological charge and the characterization of the bundle class in two or three dimensions. We also explain why a generalization to four dimensions requires extra ingredients.

The homotopy cutoff takes the gauge field induced by restricting the parallel transport of the Levi-Civita connection on the round sphere to act on unit vectors and forgets all the information in it except for \eqref{PTedges} and \eqref{PTfaces}. 
This is an example of the evaluation of the homotopy cutoff map 
\[
c_X: {\cal A}^{\infty}_{S^2 , SO(2)} \to {\cal A}^{HL}_{X_\ast , SO(2)}
\]
from the space of smooth $SO(2)$ connections over $S^2$ to the space of gauge fields on the homotopy lattice. $X_\ast$ defines a homotopy lattice for $S^2$, that considers paths on $X_1$ and homotopies among those paths restricted to $X_2$ (considered only up to homotopy), and in general homotopies among k-homotopies of paths restricted to $X_{k+1}$.

\section{HLGFs as higher parallel transport maps on the homotopy lattice}
\label{AbstactHLGFs}

%
%
%
%
%

Firstly in this section, we define Homotopy Lattice Gauge Fields (HLGFs) at an abstract level. 
Secondly, we use a trivialization on the fibers over points in the discrete set $X_0$ to provide a useful description of HLGFs. 
Finally, we give a set of generators of k-globes that provide a set of elementary building blocks for HLGFs. Since  generating 1-globes are links in $X_1$, we will see that elementary data associated with 1-dimensional objects by an HLGF is exactly standard lattice gauge field data.

\subsection{Prelude to nonabelian algebraic topology}\label{Globes+ThinHomot}


\begin{remark}\label{remark}
	It turns out that the conventions used in differential geometry and in physics follow the convention of composing curves from right to left. This forces us to adhere to this convention in our short introduction to the tools of nonabelian algebraic topology. In references on nonabelian algebraic topology usual conventions include composing from left to right. In \cite{Orendain:2023tly} we used left to right conventions but arrived at awkward conventions in which $G$-bundles had a global left action and (after local trivialization) parallel transport along composite curves could be described by right multiplication in the group. 
\end{remark}


{\em Filtered spaces and filtered maps}. 
Consider $X$ a space. 
Here our focus is on finite dimensional manifolds possibly with boundary and corners, 
which are equipped with a nested collection of subsets $X_0 \subseteq X_1 \subseteq \ldots$ such that $X$ is the union of the subsets. Such a space is called a filtered space and is denoted as $X_\ast$; the collection of nested subsets is called a filtration. 
A concrete example was given in Section \ref{PrelimExample}. 
Filtered spaces together with maps $f_\ast : X_\ast \to Y_\ast$ among them, which preserve each level of their filtrations ($f(X_k) \subseteq Y_k$ for all $k$), are organized into the category FTOP.

We will use two kinds of filtrations: Skeletal filtrations, where $X_k$ is the k-dimensional skeleton of a cellular decomposition, and trivial filtrations, where $X_k = X$ for all $k$. Filtered homotopy relative to vertices in the case of a skeletal filtration becomes thin homotopy rel. vertices, which we need to use in the base space, and in the case of a trivial filtration it becomes standard homotopy rel. vertices, which we use for the gauge group. 

{\em Skeletal filtrations}. 
First we briefly describe the homotopy theory resulting from skeletal filtrations. Our space $X$ will have a triangulation or a cellular decomposition (by a CW complex, as would be the case of a cubiculation of the n-torus by a ``Cartesian cubiculation'' with periodic boundary conditions), and $X_k$ will denote its k-skeleton. Then for k smaller or equal to the dimension of the base $\dim X_k = k$, and, by definition, $X_k =X$ for all k greater or equal to the dimension of the base. 

\begin{figure}[ht] \centering
    \includegraphics{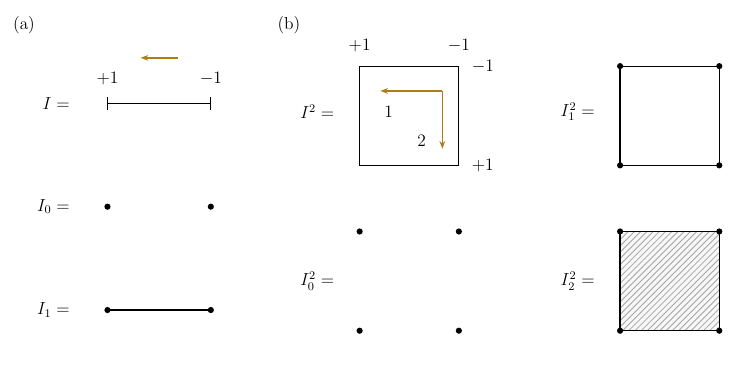}
    \caption{}
    \label{FigCubes}
\end{figure}

An example of a space with a skeletal filtration is $I^n$ the standard n-cube $[-1,1]^n$. The filtration comes from the cellular decomposition in which the k-cells are the closed k-dimensional faces of $I^n$. See Figure \ref{FigCubes} for standard 1 and 2 cubes with our convention for how to picture their standard parametrization. 

\begin{figure}[ht] \centering
    \includegraphics{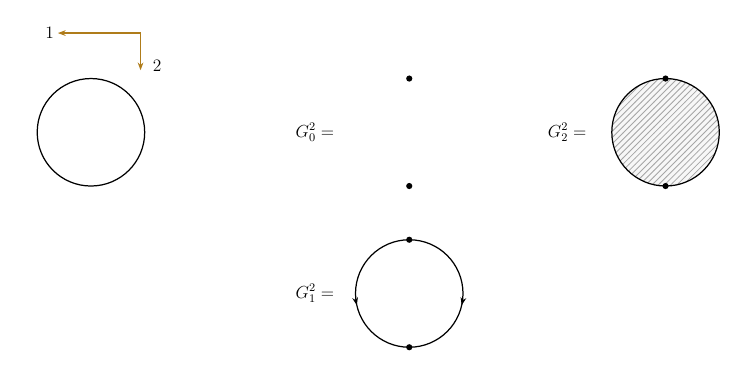}
    \caption{}
    \label{FigGlobes}
\end{figure}

Another family of examples that we will use is the family of standard n-globes. Our conventions are based on those of \cite{brown2007new}; the difference is motivated by agreeing with the conventions of differential geometry as stated in Remark \ref{remark}. In this paragraph, the difference will show up only in the picture. The standard n-globe $G^n$ is the n-dimensional disc 
$G^n = \{ x \in {\mathbb R}^n | \ || x || \leq 1 \}$ with the skeletal filtration inherited from the cellular decomposition 
$G^n = e^0_- \cup e^0_+ \cup e^1_- \cup e^1_+ \cup \ldots e^{n-1}_- \cup e^{n-1}_+ \cup e^n$ with one n-cell $e^n = Int (G^n)$ and for $0 \leq k < n$ two k-dimensional cells $e^k_+ , e^k_-$ contained in the intersection of $\partial G^n = S^{n-1}$ and the (k+1)-dimensional subspace determined by $x_j = 0$ for every $j < n-k$. Cell $e^k_\pm$ satisfies $\pm x_{n-k} > 0$. See Figure \ref{FigGlobes}. 

Now we give examples of filtered maps. They will be relevant in the algebraic structures defined below (see Figure \ref{D,E,G}) 
\[
D_i^\alpha: I^k \to I^{k+1} \quad , \quad 
(x_1, \ldots , x_k) \overset{D_i^\alpha}{\longrightarrow} (x_1, \ldots , x_{i-1}, \alpha 1 , x_i, \ldots , x_k) , 
\]
\[
E_i: I^{k+1} \to I^k \quad , \quad 
(x_1, \ldots , x_{k+1}) \overset{E_i}{\longrightarrow} (x_1, \ldots , x_{i-1}, x_{i+1}, \ldots , x_{k+1}) , 
\]
\[
G_i: I^{k+1} \to I^k \quad , \quad 
(x_1, \ldots , x_{k+1}) \overset{G_i}{\longrightarrow} (x_1, \ldots , \mbox{max} (x_i, x_{i+1}), \ldots , x_{k+1}) . 
\]

\begin{figure}[ht] \centering
    \includegraphics{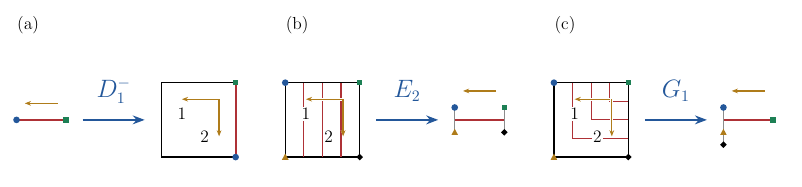}
    \caption{}
    \label{D,E,G}
\end{figure}

\begin{figure}[ht] \centering
    \includegraphics{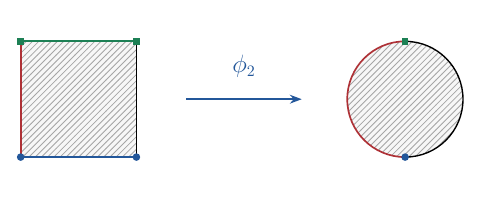}
    \caption{}
    \label{CubosAglobos}
\end{figure}

In \cite{brown2007new} a collection of filtered maps  
\[
\phi_n: I^n \to G^n  , 
\]
for all $n\geq 1$ are defined inductively. They satisfy: (1) $\phi_1 (x) = x$. (2) If we write $x=(t,y) \in I \times I^{n-1}$ then $\phi_n(t,y) = (t \sqrt{1- ||\phi_{n-1}(y)||^2}, \phi_{n-1}(y))$. (3) $\phi_n(\partial I^n) = \partial G^n$. (4) $\phi_n$ sends the $(n-1)$ dimensional face of $I^n$ transversal to direction $k \geq 1$ in ${\mathbb R}^n$ and with $\pm x^k > 0$ to the $(n-k)$ dimensional cell $e^{n-k}_\pm$. See Figure \ref{CubosAglobos} for $n=2$. 

A singular 1-globe (which is the same as a singular 1-cube) in a filtered space $X_\ast$ is a parametrized curve in $X_\ast$; at a more precise level, it is a filtered map $c: [-1,1] \to X_\ast$. Thus, it sends $[-1,1]_0$ to $X_0$ and $[-1,1]_1 = [-1,1]$ to $X_1$. Notice that this is not the same as sending vertices to vertices and edges to edges because the image of $[-1,1]$ needs to lie in $X_1$, but is not restricted to be a single edge. 
In Section \ref{PrelimExample} each edge $\Gamma_{ij}$ is a singular 1-globe, and also the Greenwich meridian running from north to south and the equator running from west to east and starting at longitude $\lambda = 0$ are singular 1-globes (see Figure \ref{TriangS2+globes}). 
The space of singular 1-globes in $X_\ast$ is denoted by $R^\bigcirc_1(X_\ast)$; previously we had called it $\tilde{P}(X_1)$. 

Similarly, a singular k-globe in $X_\ast$ is a filtered map $c: G^k \to X_\ast$, and a k-cube in $X_\ast$ is a filtered map $c: I^k \to X_\ast$. The spaces of singular k-globes and k-cubes in $X_\ast$ are denoted by $R^\bigcirc_k(X_\ast)$ and $R^\square_k(X_\ast)$ respectively. 
Each face $\Gamma_{ijk}$ in the example of Section \ref{PrelimExample} is a singular 2-globe. In \cite{Orendain:2023tly} we give an explicit map from $[0,1]^2$ to a triangulation with the images of $\{ 0 \} \times [0,1]$ and $\{ 1 \} \times [0,1]$ equal respectively to the highest and second highest vertices of a face, as shown in Figure \ref{TriangS2+globes}. (That map must be adapted to the conventions that we are now using with $I = [-1,1]$.) It is easy to see that this type of degenerate singular 2-cubes can also be seen as singular 2-globes.

Singular k-globes and k-cubes are parametrized k-dimensional subspaces of $X_\ast$ (possibly with a degenerate image) fitting into the skeletal filtration. Their special shape (globes or cubes) will let us talk about homotopies among them, and precisely describe the gluing of those homotopies.

{\em Cubical sets with ``connections''}. 
Now we will introduce $\partial_i^\alpha , \epsilon_i , \Gamma_i$ algebraic operations in $R^\square(X_\ast)$ that will let us talk about homotopies among singular cubes. 

Face maps: Given $c \in R^\square_k(X_\ast)$, we define $\partial_i^\alpha c \in R^\square_{k-1}(X_\ast)$ (for any $1 \leq i \leq k$ and $\alpha \in \{ -, + \}$) as $\partial_i^\alpha c (x_1, \ldots x_{k-1}) = c \circ D_i^\alpha (x_1, \ldots x_{k-1})$. As its name suggests, $\partial_i^\alpha c$ is the $\alpha$-face in direction $i$ of $c$. 
In our preliminary example of Section \ref{PrelimExample}, we see that $\partial_1^- \Gamma_{ij} = v_j$, $\partial_1^+ \Gamma_{ij} = v_i$. Also, at the pictorial level, 
we can see that $\partial_1^- \Gamma_{ijk} = \Gamma_{jk}$. In order to describe $\partial_2^\alpha \Gamma_{ijk}$ we need notation for degenerate singular cubes that we will give in a moment, and in order to describe $\partial_1^+ \Gamma_{ijk}$ we need notation for gluing that we will see in a few paragraphs. 

Degeneracy maps: Given $c \in R^\square_k(X_\ast)$, we define $\epsilon_i c \in R^\square_{k+1}(X_\ast)$ by \\
$\epsilon_i c (x_1, \ldots x_{k+1}) = c \circ E_i (x_1, \ldots x_{k+1})$. As its name suggests, $\epsilon_i c$ is a $(k+1)$ dimensional singular cube with the same image as $c$ which is degenerate in direction $i$. Notice that we can compose degeneracy maps to construct singular cubes with a higher degree of degeneracy. We see from Figure \ref{Gamma{ijk}As2globes} that in our preliminary example of Section \ref{PrelimExample} $\partial_2^- \Gamma_{ijk}= \epsilon_1 v_k$, $\partial_2^+ \Gamma_{ijk}= \epsilon_1 v_j$.

Connection maps: Given $c \in R^\square_k(X_\ast)$, we define $\Gamma_i c \in R^\square_{k+1}(X_\ast)$ by \\$\Gamma_i c (x_1, \ldots x_{k+1}) = c \circ G_i (x_1, \ldots x_{k+1})$. This map produces degenerate singular cubes, but the degeneracy is associated with the pair of directions $(i, i+1)$. It is because we have these maps that cubical gluing rules do not have a cartesian flavor; see Figure \ref{CubiculatedCUBE} below. 

It is not difficult to verify that 
the operations $(\partial_i^\alpha , \epsilon_i , \Gamma_i)$ defined on $R^\square (X_\ast)$ 
satisfy the relations of a ``cubical set with connections'' listed in Appendix \ref{Appendix};
here are some of mentioned relations: $\partial_i^\alpha \epsilon_j = id$ if $i=j$, $\partial_i^\alpha \epsilon_j = \epsilon_{j-1} \partial_i^\alpha$ if $i<j$, $\partial_i^\alpha \epsilon_j = \epsilon_j \partial_{i-1}^\alpha$ if $i>j$.


{\em Globular sets}. 
The set of singular globes $R^\bigcirc(X_\ast)$ admits an algebraic structure for talking about homotopies among singular globes. Below we will introduce operations $d_i^\alpha , s_{i,n}$ on $R^\bigcirc(X_\ast)$ satisfying the appropriate set of relations, giving it the structure of a globular set. 

The collection of maps $\phi$ send globes to cubes by pullback 
\[
\phi^\ast : R^\bigcirc (X_\ast) \to R^\square (X_\ast) . 
\]
Clearly, this map trivially sends 0-globes into 0-cubes and 1-globes into 1-cubes. Notice that for $n\geq 2$ the map is not surjective: its image $\phi^\ast_n (R^\bigcirc_n (X_\ast))$ is the set of singular cubes $c \in R^\square_n (X_\ast)$ such that 
$\partial_i^\alpha c \in \mbox{Im} \epsilon_1^{i-1}$ for $i\in \{ 2, \ldots , n \}$. 

It is simple to verify that the operations $\partial_1^\alpha , \epsilon_1$ preserve $\phi^\ast_n (R^\bigcirc_n (X_\ast))$. Using those operations we will define globular operations $d_i^\alpha , s_{i,n}$ in $R^\bigcirc (X_\ast)$ that make it into a globular set. Recall that $G^n$ has two $k$-dimensional cells for $k<n$; one is called the source cell of that dimension, and the other one is called the target cell of that dimension. 

Our preliminary example of Section \ref{PrelimExample} can be made precise in a simple way using globes. The singular 0-globes are the vertices in $X_0$. The singular 1-globes $\Gamma_{ij}$ shown in Figure \ref{TriangS2+globes} are continuous maps $\Gamma_{ij}: [-1, 1] \to X_1$ with $\Gamma_{ij}(-1) = v_j$ and $\Gamma_{ij}(1) = v_i$. The singular 2-globes $\Gamma_{ijk}$ are the continuous maps $\Gamma_{ijk}: G^2 \to X_2$ shown in Figure \ref{Gamma{ijk}As2globes}. 

\begin{figure}[ht] \centering
    \includegraphics{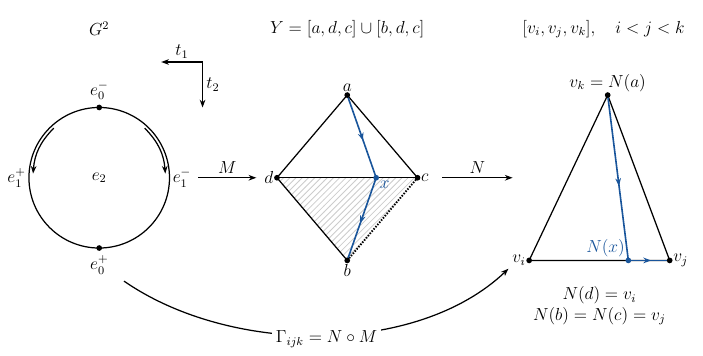}
    \caption{A homeomorphism $G^2 \to Y$, identifying the 2-disk with the 2-Rhombus presented as the gluing of two 2-simplices, lets us prescribe $\Gamma_{ijk}$ as the gluing of two maps that are affine in each simplex. The blue curve shows how $\Gamma_{ijk}$ models a homotopy of curves. From the picture we can read the face maps given below.}
    \label{Gamma{ijk}As2globes}
\end{figure}

Face maps: 
Given 
$c \in R^\bigcirc_k (X_\ast)$, for $i<k$ we define 
$d_i^\alpha c \in R^\bigcirc_i (X_\ast)$ as $c|_{e_i^\alpha}$, or equivalently as 
$d_i^\alpha c = (\phi^\ast)^{-1} (\partial_1^\alpha)^{k-i} (\phi^\ast c)$. 
They are the $i$-dimensional source and target faces of the globe. 
We will use again the preliminary example of Section \ref{PrelimExample}, but this time we will use the notation for globes. 
$d_1^- \Gamma_{ijk} = \Gamma_{jk}$, $d_0^- \Gamma_{ijk} = v_k$, $d_0^+ \Gamma_{ijk} = v_j$ (see Figure \ref{Gamma{ijk}As2globes}). $d_1^+ \Gamma_{ijk}$ will be given below. 

Degeneracy maps: 
Given 
$c \in R^\bigcirc_i (X_\ast)$, for $k>i$ we define 
$s_{i,k} c \in R^\bigcirc_k (X_\ast)$ as 
$s_{i,k} c = (\phi^\ast)^{-1} (\epsilon_1)^{k-i} (\phi^\ast c)$. 
The idea is simple, we need to extend $c$ from being defined in dimension $i$ to a higher dimension, and we do it one step at a time: Recall that $G^n \subset {\mathbb R}^n$. Consider the natural embedding of ${\mathbb R}^i$ into ${\mathbb R}^{i+1}$ at the $x^1=0$ subspace; using that embedding we can extend maps from $G^i \subset {\mathbb R}^i \subset {\mathbb R}^{i+1}$ to $G^{i+1}$ as maps that are independent of the $x^1$ coordinate. In order to bring $c$ from $R^\bigcirc_i (X_\ast)$ to $R^\bigcirc_k (X_\ast)$, the described extension has to be performed $k-i$ times. 

These maps follow the ``globular relations'': 
\begin{eqnarray*}
&&d_i^\alpha d_j^\beta = d_i^\alpha \ \mbox{ for } \ i<j , \ \ \ \ 
s_{j,k} s_{i,j} = s_{i,k} \ \mbox{ for } \ i<j , \ \ \ \ 
d_j^\alpha s_{i,k} = id \ \mbox{ for } \ i=j , \\ 
&&d_j^\alpha s_{i,k} = d_j^\alpha \ \mbox{ for } \ j<i , \ \ \ \ 
d_j^\alpha s_{i,k} = s_{i,j} \ \mbox{ for } \ j>i . 
\end{eqnarray*}

\begin{figure}[ht] \centering
    \includegraphics{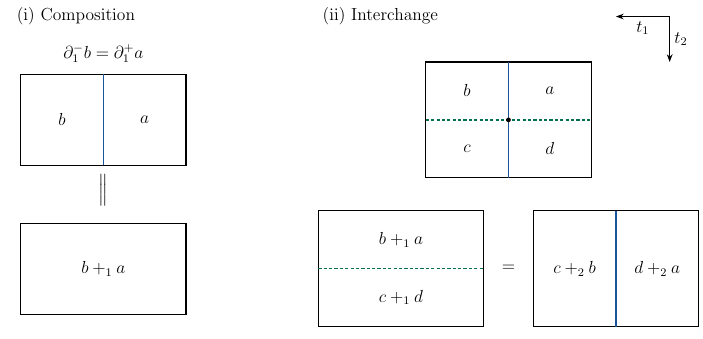}
    \caption{Panel (i) gives the pictorial image of cubical sum. Panel (ii) is the picture corresponding to the interchange rule: $(c+_1 d) +_2 (b+_1 a) = (c+_2 b)+_1 (d+_2 a)$. Solid lines indicate gluing in direction 1, and dashed lines indicate gluing in direction 2.}
    \label{SumsInterchangeRule}
\end{figure}

{\em Higher dimensional groupoid structures.} 
Now we are ready to glue homotopy maps. Here is where our convention differs from the usual convention followed in category theory treatments. The reason for our choice is given in Remark \ref{remark}. 
We describe the structure first for cubical homotopies. The operations $+_i, -_i$ on $R^\square (X_\ast)$ are defined extending the usual definitions of concatenation of curves and direction reversal of curves to higher dimensions. 
Consider $a,b \in R^\square_k (X_\ast)$ such that $\partial_i^+ a = \partial_i^- b$, then we define $b +_i a \in R^\square_k (X_\ast)$ by a simple adjustment that makes the $i$-th coordinate of the parameter space run twice as fast and use $a$ followed by $b$. 
Figure \ref{SumsInterchangeRule} illustrates gluing in two different directions for $k=2$ and the all important ``interchange rule''. 

The operation $-_i a \in R^\square_k (X_\ast)$ is defined by following the $i$th coordinate of parameter space in reversed direction. 

These operations $(+_i, -_i)$ together with the cubical operations and the connection, let us compose (or glue) homotopies among cubes in a flexible manner. As an example, Figure \ref{CubiculatedCUBE} shows the diagram corresponding to the building instructions for a cellular decomposition of $S^2$ as the boundary of a 3-dimensional cube. 

\begin{figure}[ht] \centering
    \includegraphics{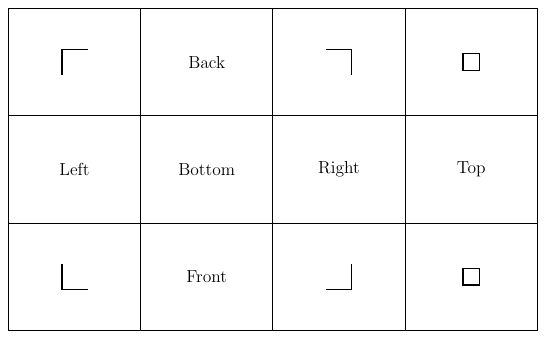}
    \caption{The diagram shows 2-cubes (or squares) glued together using operations $+_1$ and $+_2$. The interchange rule implies that we do not have to specify the order of the gluing operations. There are 6 non-degenerate squares, four degenerate squares and two doubly degenerate squares. The doubly degenerate squares are labeled by a small square. The degenerate square labeled by an L corresponds to a connection acting on a 1-dimensional face of a square, $L = \Gamma_1 (\partial_1^+ front)$. The result is that the square L identifies the edges $\partial_2^+ left$ and $\partial_1^+ front$. The reflected versions of L are obtained using $\Gamma_1$ together with $-_1$ and $-_2$.}
    \label{CubiculatedCUBE}
\end{figure}

At this point, we do not yet have a higher dimensional groupoid structure. This structure will emerge after taking a quotient by thin homotopy. 

\begin{figure}[ht] \centering
    \includegraphics{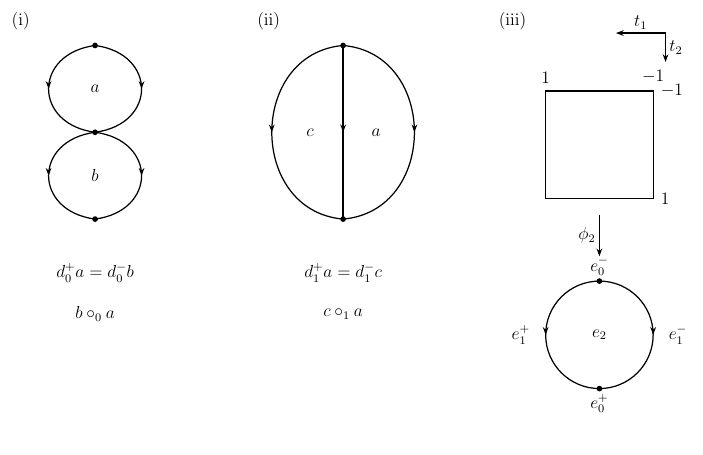}
    \caption{The $+_2$ operation for cubes induces the $\circ_0$ operation for 2-globes (gluing along 0-dimensional faces), and $+_1$ induces $\circ_1$ for 2-globes. For k-globes, the cubical $+_i$ induces the globular $\circ_{k-i}$, gluing along $(k-i)$-dimensional cells.}
    \label{+i}
\end{figure}

Relation $\phi^\ast : R^\bigcirc (X_\ast) \to R^\square (X_\ast)$ shows us that we can consider singular globes as particular cases of singular cubes. The gluing and direction changing operations $+_i, -_i$ preserve the subset $\phi^\ast (R^\bigcirc (X_\ast)) \subset R^\square (X_\ast)$, inducing corresponding gluing operations for globes. 
The picture for gluing globes is shown in Figure \ref{+i}. 

It could appear that gluing singular globes along higher codimensional cells would necessarily lead to new singular globes in which the gluing region is special. We will show, with an example, that after taking thin homotopy equivalence into account, this is not the case. 


{\em Thin homotopy}. 
In gauge theories, talking about equivalence of curves by thin homotopy is essential. Parallel transport as described in the previous section acting on two curves $c_1, c_2$ sharing their source and target points may differ even if $c_1$ and $c_2$ are homotopic rel. endpoints. Only in the case of flat connections parallel transport is a function of homotopy classes of curves relative to having fixed endpoints. On the other hand, consider the case of the constant curve $id_x$ with $x \in X_0$, and the curve $c^{-1} \circ c$ for any curve $c$ with $c(-1)=x$; this pair of curves are related by a homotopy retracting the curve $c^{-1} \circ c$ to $x$ within the image of curve. This is an example of a pair of curves related by thin homotopy. 
It turns out that for any $G$-connection $\omega$ in $M$, we have $PT_\omega(c^{-1} \circ c) = PT_\omega(id_x) = id_{{\cal F}_x}$. 
In general, for any connection and any two curves $c, c'$ with the same endpoints and related by thin homotopy relative to their endpoints we have $PT_\omega(c) = PT_\omega(c')$. 
A clearly related phenomenon is that the space of singular curves modulo thin homotopy relative to endpoints becomes a groupoid. 

In the context of a filtered space with a skeletal filtration, two singular k-cubes $c, d \in R^\square_k (X_\ast)$ such that 
$c|_{I^k_0} = d|_{I^k_0}$ are declared to be thin homotopic relative to vertices if and only if there is a 
continuous map $h: I^k \times [0,1] \to X_\ast$ with $h(v,t) = c(v) = d(v)$ for all $v\in I^k_0$ such that $h (I^k_j \times [0,1]) \subset X_j$ for all $j\leq k$ and $h(x,0) = c(x)$, $h(x,1) = d(x)$.


The set of classes of singular cubes by thin homotopy 
\[
\rho^\square X_\ast = R^\square X_\ast / \sim_{thin}
\] 
inherits the set of operations $(\partial_i^\alpha , \epsilon_i , \Gamma_i , +_i , -_i)$, following the set of relations listed in Appendix \ref{Appendix}, that make it into an $\omega$-groupoid. Its elements, singular cubes up to thin homotopy, are called cubes. 

Apart from the advantage of using a shorter name, considering cubes is motivated by parallel transport. Additionally, the gluing of cubes has a better behavior than gluing before considering equivalence classes by thin homotopy rel. vertices. Recall the preliminary example of Section \ref{PrelimExample}; it is not difficult to see that, using cubical conventions, 
$c_{\frac{2\pi}{3}} \sim_{thin} (-_2 \Gamma_{134}) +_2 \Gamma_{135}$, where all the elements of the equation are defined in Section \ref{PrelimExample}. 

The resulting category of $\omega$-groupoids is complete and cocomplete. This property is inherited by the category of infinity groupoids, that we will present below, and this will be crucial when we deal with the space of fields in the second part of this series.


We return to globes. Again we can consider the preliminary example of Section \ref{PrelimExample}. Recall that we showed explicitly how the 1-simplices of paths $\Gamma_{ijk}$ are singular 2-globes; that now we consider up to thin homotopy and call them just 2-globes. We can use globular notation to see that $c_{\frac{2\pi}{3}} \sim_{thin} (-_0 \Gamma_{134}) \circ_0 \Gamma_{135}$. This shows that gluing globes along higher codimensional cells (e.g. using $\circ_0$) does not always lead to glued regions showing their seams; in other words, having distinguishable points associated with the place where the gluing took place as it could appear from Figure \ref{+i}. 

\begin{figure}[ht] \centering 
    \includegraphics{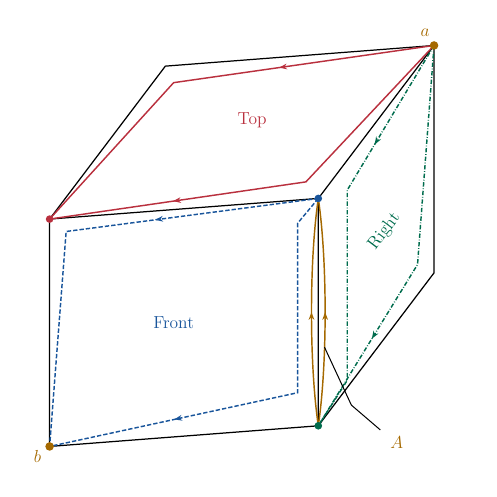}
    \caption{A cellular decomposition of $S^2$ with square cells. Each square is associated with a globe with the structure and names shown in the picture. Also one degenerate 2-globe, named $A$, is shown.}
    \label{BoundaryOfCubeWithGlobes}
\end{figure}

As another example of how globes can be glued, consider the 
construction of the 
boundary of the 3-cube using globes. We first assign 2-globes to the squares in the boundary as shown in Figure \ref{BoundaryOfCubeWithGlobes}. Degenerate 2-globes are also necessary.

A globe $G$ covering $S^2$ once can be constructed gluing three globular pieces, named $FR, BB, LT$ for Right-Front, Bottom-Back and Left-Top respectively, 
\[
FR \simeq_{thin} Front \circ_0 A \circ_0 Right . 
\]
Notice that $d_0^- FR = a$ and $d_0^+ FR = b$. With similar appropriate definitions for $BB$ and $LT$, they will all have the same 0-dimensional source and target, and also share pairwise their 1-dimensional source and target. Then 
\[
G = LT \circ_1 BB \circ_1 FR
\]
is a 2-globe covering $S^2$ once as promised. The reader is invited to make this construction explicit.

The set of classes of singular globes by thin homotopy 
\[
\rho^\bigcirc X_\ast = R^\bigcirc X_\ast / \sim_{thin}
\] 
inherits the set of operations $(d_i^\alpha , s_{i,k} , \circ_i , -_i)$, following the globular relations stated above and relations stating compatibility with the gluing operations (stated below), making it into a strict $\infty$-groupoid. Its elements, singular globes up to thin homotopy, are called globes. %
The relations stating compatibility of the gluing operations with the globular operations and internal compatibility of the gluing operations are as follows. For $a,b\in\rho^\bigcirc_n X_*$ and $0\leq j<n$, the composite $a\circ_j b$ is defined precisely when $d_j^-a=d_j^+b$, meaning $b$ followed by $a$.

For $\alpha\in\{-,+\}$, faces of dimension greater than $j$ preserve composition:
\[
d_i^\alpha(a\circ_j b)
=
d_i^\alpha a\circ_j d_i^\alpha b,
\qquad 0\leq j<i<n.
\]
At dimension $j$, the source and target satisfy
\[
d_j^-(a\circ_j b)=d_j^-b,
\qquad
d_j^+(a\circ_j b)=d_j^+a.
\]
The lower-dimensional faces satisfy
\[
d_i^\alpha(a\circ_j b)=d_i^\alpha a=d_i^\alpha b,
\qquad 0\leq i<j<n.
\]
Degeneracy maps preserve composition:
\[
s_{n,k}(a\circ_j b)
=
s_{n,k}a\circ_j s_{n,k}b,
\qquad 0\leq j<n<k.
\]
Finally, for $a,b,c,d\in\rho^\bigcirc_n X_*$, the interchange law is
\[
(a\circ_j b)\circ_i(c\circ_j d)
=
(a\circ_i c)\circ_j(b\circ_i d),
\qquad 0\leq i,j<n,\quad i\neq j,
\]
whenever the indicated composites are defined.

%

{\em Trivial filtrations}. 
Our physical motivation needs a cutoff for quantum field theory, making the space of histories (also called the space of field configurations) finite dimensional. We achieve this goal introducing a filtration in spacetime's path groupoid and considering paths up to filtered homotopy rel. vertices. On the other hand, the gauge group playing the central role in the target space of our HLGFs cannot have a filtration where $G_0 \neq G$. The reason is that physics calls for $G_0 = G$ and nestedness forces $G_k = G$ for all $k$. 
We could work with a filtration using a discrete subgroup $G_0$, but we would have to live with that structure without the possibility of getting rid of that cutoff by using ``finer and finer discrete subgroups''. In the 80s discrete subgroups were used in simulations within lattice gauge theory, but as soon as the computer power allowed them to abandon that drastic simplification the use of discrete subgroups became a historical curiosity. 

A filtration in $G$ with $G_0$ a discrete subgroup and $G_i = G$ for $i > 0$ could be interesting for the calculation of topological invariants; we do not pursue this route in this work. 

Below we briefly describe the fundamental higher homotopy groupoid $\rho^\bigcirc G$ in the case where the filtration in the space $G$ is the trivial filtration; i.e., $G_i = G$ for all $i$. We omit the subscript $\ast$ to show in the notation that the trivial filtration is used. 

It is unfortunate that the standard notation uses $G$ for the gauge group and the standard globular shape for k-globes is also denoted using the same letter. We will keep the standard notation. The context should help the reader avoid confusion; also the standard k-dimensional globular shape $G^k$ has a superscript and the gauge group doesn't. 

k-globes in $\rho^\bigcirc G$ are equivalence classes of filtered maps from $G^k$ to $G$ relative to vertices. The set of vertices in the domain $G^k_0$ has two elements $e_0^-$ and $e_0^+$. Their images $c(e_0^-), c(e_0^+) \in G$ are arbitrary points in $G_0=G$. Thus, k-globes are filtered homotopy classes of maps $c: G^k \to G$ relative to having the points $c(e_0^-), c(e_0^+) \in G$ fixed during the homotopy. Since we are using the trivial filtration, filtered homotopy rel. vertices becomes usual homotopy rel. vertices. 

In the case of 1-globes with $c(e_0^-) = c(e_0^+)$, homotopy rel. vertices becomes the standard notion of homotopy of closed based curves rel. their base point. In other words, $\rho_1^\bigcirc G$ has subgroups of the type $\pi_1(G, \ast)$ for $\ast$ being any point in $G$. 

For the moment forget that our $G$ is a Lie group. Could we capture the information about the second homotopy group of a space $\pi_2(G, \ast)$ from $\rho^\bigcirc G$ truncated at level two? The candidate singular globes containing that information are 2-singular globes with $\tilde{c}(e_0^-) = \tilde{c}(e_0^+) = \ast$ and $\tilde{c}(e_1^-) = \tilde{c}(e_1^+)= s_{0,1} \ast$. We add a tilde to remind us that they are not 2-globes because we have not considered their equivalence class up to filtered homotopy rel. vertices. $\tilde{c}$ could be considered as a based map from $(S^2, \ast)$ to $(G, \ast)$, and we could consider $<<\tilde{c}>>$ its homotopy class among such maps. We remark that $<<\tilde{c}>> \in \pi_2(G, \ast)$ is not the same kind of equivalence class as $c=[\tilde{c}] \in \rho_2^\bigcirc G$. 
The big difference is that during the homotopies defining $c$ the image of $\tilde{c}(e_1^-)$ and $\tilde{c}(e_1^+)$ can be different from the constant curve $s_1 \ast$. On the other hand, the definition of $<<\tilde{c}>>$ as the equivalence class of maps from $G^2$ requires that during the homotopy the image of the complete boundary of $G^2$ remains to be the base point. 

The same phenomenon happens for higher dimensional globes: $c=[\tilde{c}]$ is the same type of equivalence class as $<<\tilde{c}>>$ only for 1-globes. 

Since $\pi_2 G$ is trivial for all Lie groups, working with $\rho^\bigcirc G$ fails to give us information about the homotopy type of $G$ starting at the third homotopy group. In Section \ref{TopChargeSection} we will see that this is the reason that a HLGF determines a $G$ bundle for base spaces of dimension two and three, but in general not for higher dimensional base spaces.

The category of strict $\infty$-groupoids turns out to be equivalent to the category of cubical $\omega$-groupoids with connections.

\subsection{Abstract HLGFs and the higher homotopy Atiyah groupoid}

In section \ref{Sec2} we described gauge fields in the continuum as smooth parallel transport maps. In order to state the properties of parallel transport maps in convenient algebraic terms, we presented the Atiyah groupoid: 
Parallel transport maps are sections of the canonical projection from the Atiyah groupoid to the path groupoid. 
Later, we motivated that HLGFs should be parallel transport maps along globes transporting relative homotopies (of initial conditions) in the fiber over the source point to relative homotopies (of final conditions) in the fiber over the target point. Here we will formalize these ideas. 

We will start describing in detail what $At(X_1; G)$ the Atiyah groupoid over $X_1$ is. This is the appropriate groupoid to talk about parallel transport on $X_1$. Such parallel transport maps correspond to standard lattice gauge fields. In section \ref{ELGFs} we will describe these fields in terms of generators; the resulting picture will be recognizable as a standard gauge field, where a gauge field is given in terms of data associated with the links of $X_1$. 

Consider ${\cal F}$ a torsor for the Lie group $G$; that is, it is a smooth manifold with a transitive and free action of $G$. Left and right actions will both play different roles below. ${\cal F}$ will be the typical fiber of the $G$-bundles that we will construct. 

The pair groupoid $Pair (X_0)$ has $X_0$ as set of objects and its set of morphisms is the set of ordered pairs of objects. 
Source and target maps are obvious, and composition of composable morphisms is also clear. 

$At(X_0; G)$ the Atiyah groupoid over $Pair (X_0)$ is a modification of $Pair (X_0)$ with set of objects and set of morphisms corresponding to fibers over points in $X_0$ and $G$ equivariant maps among them. 
In detail, it is convenient to define 
objects to be pairs $(x, {\cal F}_x)$ for some $x \in X_0$, and  
morphisms to be pairs $((x, y), M_{(x,y)})$ where the first entry is an ordered pair of points in $X_0$ and the second one is a $G$-equivariant map $M_{(x,y)} : {\cal F}_x \to {\cal F}_y$. 
The source map is $s(((x, y), M_{(x,y)})) = (x, {\cal F}_x)$ and the target map is $t(((x, y), M_{(x,y)})) = (y, {\cal F}_y)$. 
Composition of $G$ equivariant maps is another $G$ equivariant map; the composition of morphisms in $At(X_0; G)$ is a new pair where the first entry is a composition of pairs and the second one a composition of $G$ equivariant maps. 
As intended, $At(X_0; G)$ has a canonical projection to $Pair (X_0)$. 
Finally, notice that the set of $G$ equivariant maps is conveniently described by its 1 to 1 correspondence with the set $({\cal F}_x \times {\cal F}_y) / G$, where the quotient is by the diagonal right $G$ action. 
We will summarize the construction by writing 
\[
At(X_0; G)=Pair(X_0 \times {\cal F}) / G . 
\]

\begin{figure}[ht] \centering
    \includegraphics{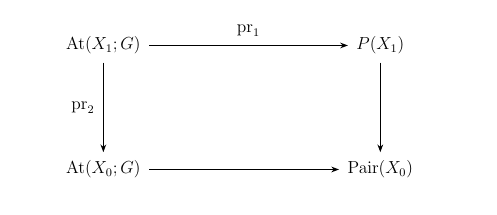}
    \caption{}
    \label{AtX_1}
\end{figure}

$At(X_1; G)$ the Atiyah groupoid over $X_1$ is the enrichment of $\rho^\bigcirc_1(X_\ast)=P(X_1)$, the path groupoid of $X_1$ (which has a canonical projection to $Pair(X_0)$), needed for considering its morphisms as parallel transport maps. Thus, it is defined by the following pullback diagram (Figure \ref{AtX_1}) 

\begin{figure}[ht] \centering
    \includegraphics{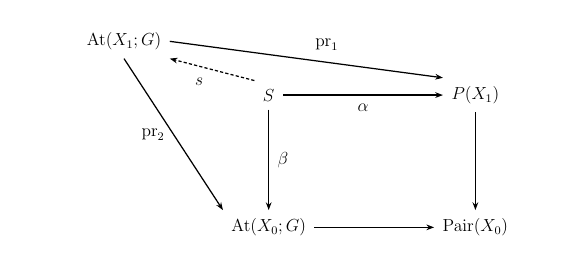}
    \caption{}
    \label{Pullback}
\end{figure}

The meaning of the previous definition ``by pullback'' is the following: We know all the corners of the diagram except for the top left corner. We also know all the arrows except for the ones starting in the top left corner. 
In the problem of finding a groupoid in the top left corner together with its arrows which make the diagram into a commuting square, our unknowns are $(At(X_1; G), pr_1, pr_2)$. 
We would like to find a solution to the problem which is minimal and sufficient. A solution of this problem is considered minimal and sufficient if and only if for any other solution $(S, \alpha , \beta)$ the arrows can be obtained by pull back from a unique map $s: S \to At(X_1; G)$ as shown below (Figure \ref{Pullback}) 

This procedure is considered a definition because, in the context of a complete and cocomplete category like the one we are using, it is known that the problem has an answer, and it is essentially unique ($At(X_1; G)$ is unique up to a unique isomorphism commuting with 
the projections to $At(X_0; G)$ and $P(X_1)$). 

An explicit description of $At(\pi)$ at a heuristic level was given in section \ref{PTsubsection}. That description is easily adapted to describe $At(X_1; G)$. The reader is invited to verify that the resulting description does satisfy the definition given above. 

Before proceeding, let us clarify a point where common language and the previous technical definitions may not match. We repeatedly mentioned the parallel transport of initial conditions on the fiber over the source of a curve in the base space; we were careful to refer to the initial conditions as ``elements being parallel transported'' to avoid calling them ``objects being parallel transported''. 
We are not interested in transporting only one initial condition, but any initial condition. 
Because of this, we may consider that the objects being transported along a curve $c$ are entire fibers ${\cal F}_{s(c)}$. 
The technical terminology used above calls entire fibers the objects. 
The parallel transport map associated with a curve is a map from one fiber to another one; using that map we can transport individual initial conditions. 
This is the picture portrayed when we define parallel transport maps as sections 
\[
P(X_1) \to At(X_1; G)
\]
of the projection $At(X_1; G) \to P(X_1)$. 

This is the picture that will be extended to construct homotopy parallel transport. 

Now, guided by the above discussion in Section \ref{HHPTcontinuum}, we proceed to construct $At^{1,\bigcirc}(X_\ast; \rho^\bigcirc G)$, the appropriate structure to describe homotopy parallel transport in the homotopy lattice, by means of a pull-back definition of the type used for $At(X_1; G)$ as illustrated in Figure \ref{AtX_1}. We need to provide a replacement for the groupoid in the top right corner of the diagram, which is appropriate to talk about homotopies of paths. A k-homotopy of paths will be modeled by a (k+1)-globe, called a k-globe of paths. The place where globes of paths live will be denoted by $\rho^{1,\bigcirc} X_\ast$; it is the same set as $\rho^\bigcirc X_\ast$ with a slight reorganization of its structure associated to thinking about (k+1)-globes as k-globes of paths. 
Also a replacement to the lower left corner of the pull-back diagram will be provided. We will denote it by $At^{1,\bigcirc}(X_0; \rho^\bigcirc G)$. 
Its structure will tell us that what can be transported are k-homotopies of initial conditions (along k-globes of paths), and that the transport is equivariant with respect to the right action of $G$ acting pointwise on homotopies of initial conditions. 

Let us give details about how $\rho^{1,\bigcirc} X_\ast$ reorganizes the structure of $\rho^\bigcirc X_\ast$. 
The product $\circ_0$ is singled out, because it talks about composition in a path-way gluing along 0-dimensional boundaries, and treated as the ``outer groupoid structure''. For that outer groupoid structure morphisms will be called arrows to help us separate this part of the structure from the ``inner strict infinity groupoid'' part of the structure talking about homotopies of paths. These two structures are coordinated by the same set of properties that held before we introduced this change of terminology. In our application to gauge fields, this terminology update is not merely cosmetic. We mentioned that parallel transport will take homotopies of paths to an appropriately $G$ equivariant map; that notion will be translated into equivariance with respect to $\rho^\bigcirc G$, which, since $G$ is a manifold and also a group, apart from its homotopy structure has a product, among k-globes inherited from the product in $G$ acting pointwise. That extra product will give rise to the ``outer groupoid product'' that will be paired to outer groupoid structure in $\rho^{1,\bigcirc} X_\ast$. It is easy to verify that this ``outer structure'' is compatible with the homotopic part of the structure in $\rho^\bigcirc G$, which will  be considered the interior part. We consider both structures together and use the notation $\rho^{1,\bigcirc} G$. 
Let us go back to finish our description of $\rho^{1,\bigcirc} X_\ast$. 
The arrows of the outer groupoid structure are k-globes of paths. For example 
$c \in \rho^\bigcirc_{k+1} X_\ast$ will be considered an arrow. 
Then $d_0^- c$ will be an object sitting on a vertex in $X_0$. Since a k-globe of paths and a j-globe of paths can only be $\circ_0$-composed if $j=k$, on any given vertex of $X_0$ we have different types of objects labeled by the level $k$. The set of objects may be identified with the discrete strict infinity groupoid $\rho^\bigcirc (X_0)_\ast$, and its restriction to the elements over any given $x \in X_0$ is $\rho^\bigcirc x_\ast$ which is the infinity groupoid formed by the objects of $\rho^{1,\bigcirc} X_\ast$ sitting over $x$, and can also be identified with the set of identity elements of the outer groupoid structure associated to $x$. We saw that $d_0^-$ plays the role of the source operation in the outer groupoid structure, likewise, $d_0^+$ plays the role of the target operation, and $-_0$ plays the role of the inverse. 
In addition, given a pair $(x,y) \in Pair (X_0)$ the set of arrows joining objects sitting over the elements of the pair has the structure of a strict infinity groupoid where all the operations on $\rho^\bigcirc X_\ast$, except for the 0th level, give the ``internal structure'' to that set of arrows. 

From what is written above, we see that the lower right corner of the diagram needs an updated version of $Pair (X_0)$, where the set of objects over any $x \in X_0$ is labeled by the level $k$, and the set of arrows over a pair is also labeled by the level. Notice that it trivially has the structure of a groupoid internal to a strict infinity groupoid. We will call it $Pair^{1,\bigcirc} (X_0)$. With this structure, it is clear that there is a morphism of groupoids internal to strict infinity groupoids 
projecting $\rho^{1,\bigcirc} X_\ast$ to $Pair^{1,\bigcirc} (X_0)$; for objects the projection preserves labels $x \in X_0$ and also preserves the levels indexed by a value of $k$. For arrows, the k-globe of paths $c$ is taken to the arrow labeled by $(x = d_0^- c, y = d_0^+ c)$ and level $k$. 

Now let us define the lower left corner. The expression  
\[
At^{1,\bigcirc}(X_0; \rho^\bigcirc G) = Pair^{1,\bigcirc}(X_0 \times \rho^\bigcirc {\cal F}) / \rho^\bigcirc G , 
\]
is a useful summary after the following clarifications: 
${\cal F}$ is  a right $G$-torsor considered as a filtered space with the trivial filtration; $G$ also has the trivial filtration. 
Its set of arrows is shown in the expression. The product, taken at each level $k$ representing pairs of sets of possible level $k$ initial/final conditions (k-homotopies in ${\cal F}$), is then considered only up to equivalence by the diagonal action of $\rho^\bigcirc_k G$ with a right action on $\rho^\bigcirc_k{\cal F}$ as appropriate for a transport map. 
An arrow is of the form $((x,y), T_k)$ with $(x,y) \in Pair (X_0)$ and $T_k : \rho^\bigcirc_k {\cal F}_x \to \rho^\bigcirc_k {\cal F}_y$ a $\rho^\bigcirc_k G$ equivariant transformation. 
Again, the set of arrows sitting over an element of $Pair (X_0)$ has the structure of a strict infinity groupoid. 
Its set of objects is in one to one correspondence with $\rho^\bigcirc (X_0)_\ast$, and a given object is naturally identified with $\rho^\bigcirc_k {\cal F}_x$ for some vertex $x \in X_0$ and some level $k$. 
Source and target maps are clear. The morphism of groupoids internal to strict infinity groupoids 
projecting $At^{1,\bigcirc}(X_0; \rho^\bigcirc G)$ to $Pair^{1,\bigcirc} (X_0)$ is also obvious; it preserves levels and pair labels.

The pull-back diagram in the category of groupoids internal to infinity groupoids defining $At^{1,\bigcirc}(X_\ast; \rho^\bigcirc G)$ is the following 

\begin{figure}[ht] \centering
    \includegraphics{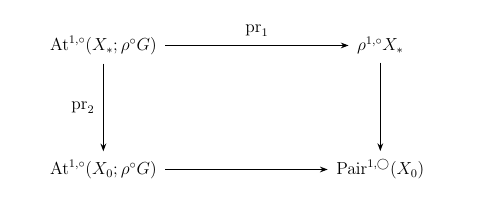}
    \label{AtX_ast}
\end{figure}

To gain a more concrete understanding, let us explicitly describe the structure of $At^{1,\bigcirc}(X_\ast; \rho^\bigcirc G)$. 
Its objects are of the form $(x; \rho^\bigcirc_k {\cal F}_x)$ for some $x \in X_0$ and for some $k\geq 0$. 
The set of arrows is in correspondence with arrows in $At^{1,\bigcirc}(X_0; \rho^\bigcirc G)$ and in $\rho^{1,\bigcirc} X_\ast$. Given a pair $(x, y)$, 
arrows are of the form $(c; T)$, where $c$ is a k-globe of paths (with $d_0^- c = x$, $d_0^+ c = y$) and $T: \rho^\bigcirc_k{\cal F}_x \to \rho^\bigcirc_k{\cal F}_y$. 

Source and target maps in $At^{1,\bigcirc}(X_\ast; \rho^\bigcirc G)$ are obvious: 
$s(c; T_k)= (x ; \rho^\bigcirc_k {\cal F}_x)$ with $x = d_0^- c$
and 
$t(c; T_k)= (y ; \rho^\bigcirc_k {\cal F}_y)$ with $y = d_0^+ c$. 
The other structural maps are obvious; for example, 
the composition of two composable arrows is $(c^2; T^2_k) \circ_0 (c^1; T^1_k) = (c^2 \circ_0 c^1; T^2_k \circ T^1_k)$. 

The projection 
$pr_1 : At^{1,\bigcirc}(X_\ast; \rho^\bigcirc G) \to \rho^{1,\bigcirc} X_\ast$ acting on objects takes $(x; \rho^\bigcirc_k {\cal F}_x)$ to 
$\rho^\bigcirc_k {x}_\ast$. 
Its action on arrows is $pr_1 (c; T_k) = c$. 
On the other hand, the projection 
$pr_2 : At^{1,\bigcirc}(X_\ast; \rho^\bigcirc G) \to At^{1,\bigcirc}(X_0; \rho^\bigcirc G)$ acting on objects takes $(x; \rho^\bigcirc_k {\cal F}_x)$ to 
$\rho^\bigcirc_k {\cal F}_x$. 
Its action on arrows is $pr_2 (c; T_k) = ((x,y), T_k)$ with $x = d_0^- c$ and $y = d_0^+ c$. 

Parallel transport maps in the homotopy lattice, abbreviated as HLGFs, correspond to sections (morphisms of groupoids internal to strict infinity groupoids) of the projection in the top row of the diagram 
\begin{equation}\label{PT}
	PT: \rho^{1,\bigcirc} X_\ast  \to At^{1,\bigcirc}(X_\ast; \rho^\bigcirc G) . 
\end{equation}

A given initial condition $u \in \rho^\bigcirc_k {\cal F}_x$ 
is transported by a k-globe of paths $c$ with $(s(c), t(c))= (x,y)$ to the final condition $v = T_k u  \in \rho^\bigcirc_k {\cal F}_y$, where $PT(c) = (c; T_k)$.

Like with ordinary fields, our gauge fields are now seen as sections. Since the non trivial information is all contained in the second entry of the image, a short hand notation, similar to the one used for ordinary fields writing $\phi(x)$ instead of $(x,\phi(x))$, is 
\[
c \overset{PT}{\longmapsto} PT(c): \rho^\bigcirc_k{\cal F}_x \to \rho^\bigcirc_k{\cal F}_y . 
\]

The following table summarizes the degree shift between globes in the base and their corresponding parallel transport data. Below $x=d_0^-c$ and $y=d_0^+c$. The last column anticipates the description after choosing a trivialization over vertices, which is introduced in the next subsection.

\begin{table}[ht]
\begin{tabular}{lccc}
\hline
\shortstack[l]{Globe $c$\\in the base}
&
\shortstack{Degree as a\\globe of paths}
&
\shortstack{Transport map\\$PT(c)$}
&
\shortstack{Trivialized value\\belongs to}
\\
\hline
Path: a $1$-globe
& $0$
& $\mathcal F_x\longrightarrow\mathcal F_y$
& $G$
\\[6pt]
\shortstack[l]{Homotopy of paths:\\a $2$-globe}
& $1$
& $\rho^\bigcirc_1\mathcal F_x
   \longrightarrow\rho^\bigcirc_1\mathcal F_y$
& $\rho^\bigcirc_1G$
\\[6pt]
\shortstack[l]{A $(k+1)$-globe,\\$k\geq0$}
& $k$
& $\rho^\bigcirc_k\mathcal F_x
   \longrightarrow\rho^\bigcirc_k\mathcal F_y$
& $\rho^\bigcirc_kG$
\\
\hline
\end{tabular}
\caption{Degrees of base globes and their transport data.
The last column uses a trivialization over vertices.}
\label{tab:HLGFdegrees}
\end{table}

\subsection{HLGFs after a trivialization over vertices}

In the differential geometric treatment of $G$ bundles and connections, a local trivialization is used to describe any particular connection with respect to the trivialization. The result is a concrete description. Concreteness comes at the price of having to take care of relating descriptions that differ only by the use of different local trivializations. In this subsection we will present a concrete description of HLGFs relying on the choice of a trivialization on the fibers over $X_0$. 

It is very important to realize that because $X_0$ is a discrete set, we can choose a trivialization on the fibers over $X_0$ without the need to work locally and deal with compatibility at chart intersections that is unavoidable when allowing a continuum set of fibers%
\footnote{In \cite{Orendain:2023tly} we presented two equivalent descriptions of HLGFs working with trivializations, one uses globes as we have done here. The other one is based on cubes instead of globes, and it was very useful for proving key properties, but it has the inconvenience that it needs local trivializations with non trivial intersections as in the usual approach in the continuum. In that context, a HLGF is a consistent collection of local HLGFs relative to local trivializations.}. 

A trivialization $\phi$ is a set of identifications of the right $G$-torsors with the group itself $\{ \phi(x): {\cal F}_x \to G \}_{x \in X_0}$ compatible with the right actions of $G$. Using those identifications, a $G$-equivariant map ${\cal F}_x \to {\cal F}_y$ can be implemented by a map $G\to G$ commuting with right translations. Then one notices that the set of those maps is in one to one relation with the set of left translations, justifying the parametrization of parallel transport maps by elements of $G$ acting by left multiplication. 
Recall that the composition of two $G$ equivariant maps is described by the corresponding group elements multiplied in a left-wise order. 
Similarly, a right 
$\rho^\bigcirc G$-equivariant map $\rho^\bigcirc {\cal F}_x \to \rho^\bigcirc {\cal F}_y$ translates into 
a right $\rho^\bigcirc G$-equivariant map $\rho^\bigcirc G \to \rho^\bigcirc G$, and all such maps can be obtained by left multiplication. Thus, relative to the trivialization, a HLGF $PT$ acting on a k-globe of paths 
$PT(c): \rho^\bigcirc_k{\cal F}_{s(c)} \to \rho^\bigcirc_k{\cal F}_{t(c)}$ is encoded by an element 
\[
PT_\phi(c) \in \rho^\bigcirc_k G 
\]
acting by left multiplication on $\rho^\bigcirc_k G$. We have a collection of these assignments for every pair of base points $(x,y)$ and every level of globes of paths. They are summarized by 
an assignment $PT_\phi$ from $\rho^{1,\bigcirc} X_\ast$ to $\rho^\bigcirc G$, where the assignment sends k-globes of paths (which is the same as $k+1$ globes in $\rho^\bigcirc X_\ast$) to k-globes in $\rho^\bigcirc G$. 

That $PT: \rho^{1,\bigcirc} X_\ast  \to At^{1,\bigcirc}(X_\ast; \rho^\bigcirc G)$ is a morphism 
must translate into some property of $PT_\phi$. The clue to understanding this property is the mismatch between the structures in the expression that we want to understand: On one side we have $\rho^{1,\bigcirc} X_\ast$ that has the structure of a groupoid internal to strict infinity groupoids as described in the previous subsection. 
On the other side, we have $\rho^\bigcirc G$ that we are using for two reasons, its group structure (inherited from the group structure of $G$) and its infinity groupoid structure whose stratification into levels 
lets us encode right $(\rho^\bigcirc G)_k$-equivariant maps $(\rho^\bigcirc {\cal F}_x)_k \to (\rho^\bigcirc {\cal F}_y)_k$ by left multiplication by elements of $(\rho^\bigcirc G)_k$. 
Since we may consider groups $G$ as groupoids $BG$ with a single object, we see that $\rho^\bigcirc BG$ can be given the structure of 
a groupoid internal to strict infinity groupoids
(the groupoid structure inherited from that of $BG$, and the strict infinity groupoid structure following from higher homotopy). We will denote the corresponding structure by $\rho^{1,\bigcirc} BG$. 
Its formal definition is $\rho^{1,\bigcirc} BG = B \rho^\bigcirc G$. 
At each level it has just one object $\ast_k$, its arrows at level $k$ are $Arrow \left( \rho^{1,\bigcirc} BG \right)_k =  \rho^\bigcirc_k G$, and they organize into a strict infinity groupoid. 

By construction, we see that a trivialization $\phi$ transforms $PT$ into a morphism of groupoids internal to strict infinity groupoids where the composition $\circ$ and inverse ${}^{-1}$ of the groupoid is in correspondence with 
the degreewise multiplication and multiplicative inverse in $\rho^{1,\bigcirc} BG$, induced by those of $BG$.  
In summary, a trivialization $\phi$ transforms $PT$ into a morphism 
\begin{equation}\label{PTphi}
	PT_\phi : \rho^{1,\bigcirc} X_\ast \to \rho^{1,\bigcirc} BG . 
\end{equation}
Conversely, with the vertex trivializations fixed, every such morphism determines a unique HLGF.

In \cite{Orendain:2023tly} we had defined HLGFs as morphisms of infinity groupoids 
\begin{equation}\label{Aphi}
	A_\phi : \rho^\bigcirc X_\ast \to \rho^\bigcirc G [-1]. 
\end{equation}
We will not give the definition of $\rho^\bigcirc G [-1]$ here, but a simple comparison makes it obvious that the definitions agree. See the Table at the end of the previous subsection.

A change to a different trivialization $\phi^2$ of the fibers over $X_0$ results in a morphism $PT_{\phi^2}$ related to $PT_{\phi^1}$ by conjugation. First, 
changing from trivialization 
$\{ \phi^1(x): {\cal F}_x \to G \}_{x \in X_0}$ to trivialization 
$\{ \phi^2(x): {\cal F}_x \to G \}_{x \in X_0}$ 
is accomplished by left multiplication by the transition functions $\{ \psi^{12}(x) \in G \}_{x \in X_0}$ on $G$ as follows: 
$\{ \phi^2 (x) = L_{\psi^{12}(x)} \circ \phi^1(x) = \psi^{12}(x) \cdot \phi^1(x) \}$. 
Then for any k-globe of paths we have 
\[
PT_{\phi^2} (c) = s_{0,k}(\psi^{12}_{t(c)}) \cdot (PT_{\phi^1} (c)) \cdot s_{0,k}(\psi^{12}_{s(c)})^{-1} . 
\]

\subsection{Gauge transformations}

In the differential geometric picture, gauge transformations are induced by vertical bundle automorphisms. A vertical bundle automorphism transforms a given connection to another one. At a local level with respect to a given local trivialization, these transformations act by conjugation as in the last formula of the previous subsection (but within the given local trivialization). 
For each given bundle, these transformations form a group defining gauge equivalence classes in the space of connections on that bundle. 
If all $G$-bundles over a base space $M$ are treated at once in the standard continuum treatment, gauge transformations form a groupoid that is a disjoint union of the vertical automorphism groups of each separate bundle (two vertical bundle automorphisms can be composed only if they act on the same bundle). 
From the algebraic point of view in the continuum picture heuristically described in Subsection \ref{HHPTcontinuum}, considering a groupoid of paths in $M$ with a discrete collection of base points, we described gauge fields on all possible $G$-bundles over a given base $M$ as groupoid homomorphisms (the heuristic counterpart in the continuum of definition \eqref{PT} in our discrete setting). 
Notice however, that HLGFs are only sensitive to the restriction of vertical bundle automorphisms to the bundle over the discrete set $X_0$, and that the resulting automorphisms over the $G$ bundle over $X_0$ do form a group. 

A vertical automorphism of the bundle over $X_0$ is characterized by a map $g:X_0 \to G$. The induced gauge transformation acting on HLGFs is $PT' = g \triangleright PT$; its evaluation on $c$ a k-globe of paths, described using a trivialization $\phi$ over $X_0$, yields 
\[
PT'_{\phi} (c) = s_{0,k}(g_{t(c)}) \cdot (PT_{\phi} (c)) \cdot s_{0,k}(g_{s(c)})^{-1} . 
\]
Considering the abstract definition of a HLGF of equation \eqref{PT}, $PT: \rho^{1,\bigcirc} X_\ast  \to At^{1,\bigcirc}(X_\ast; \rho^\bigcirc G)$, the gauge transformation $g: PT \Rightarrow PT'$ 
is a strict natural isomorphism over the identity of $\rho^{1,\bigcirc} X_\ast$. 
Let us recall that both $PT$ and $PT'$ are sections of the projection $At^{1,\bigcirc}(X_\ast; \rho^\bigcirc G) \to \rho^{1,\bigcirc} X_\ast$, and as such they send objects in $\rho^{1,\bigcirc} X_\ast$, which are degenerate k-globes over some $x \in X_0$ for some $k$, to objects $(x, \rho_k^\bigcirc{\cal F}_x)$ in $At^{1,\bigcirc}(X_\ast; \rho^\bigcirc G)$ and 
arrows in $\rho^{1,\bigcirc} X_\ast$, which are (k+1)-globes in $\rho^\bigcirc X_\ast$, 
to arrows $(c, T_c^k)$ in $At^{1,\bigcirc}(X_\ast; \rho^\bigcirc G)$ at each level $k$ and preserving the internal infinity groupoid structure and the outer groupoid structure. 
At a given level $k$, the component of the natural transformation over $x$ is $g_x: (x, \rho_k^\bigcirc{\cal F}_x) \to (x, \rho_k^\bigcirc{\cal F}_x)$ the morphism acting on $\rho_k^\bigcirc{\cal F}_x$ by left multiplication by $s_{0,k}(g_{x})$. Given any arrow $c$ in $\rho^{1,\bigcirc} X_\ast$ of level $k$ with $s(c)=x$, $t(c)=y$ the following diagram in $At^{1,\bigcirc}(X_\ast; \rho^\bigcirc G)$ commutes

\begin{center}
\begin{tikzpicture}
\matrix(m)[matrix of math nodes, row sep=4em, column sep=4em,text height=1.5ex, text depth=0.25ex]
{(x,\rho_k^\bigcirc \F_x) & (y,\rho_k^\bigcirc \F_y)\\
(x,\rho_k^\bigcirc \F_x)& (y,\rho_k^\bigcirc \F_y)\\};
\path[->,font=\scriptsize,>=angle 90]
(m-1-1) edge node [above]{$PT(c)$} (m-1-2)
        edge node [left]{$s_{0,k}(g_x)$} (m-2-1)
(m-1-2) edge node [right]{$s_{0,k}(g_y)$}(m-2-2)
(m-2-1) edge node [below]{$PT'(c)$}(m-2-2)

;
\end{tikzpicture}
\end{center}

Notice that the definition of homotopy parallel transport maps as commuting with the right action of $\rho^\bigcirc G$ on fibers over objects forces 
all vertical natural transformations 
connecting HLGFs to be determined at each level by left multiplication by $s_{0,k} g$ for some $g:X_0 \to G$. The same $g:X_0 \to G$ must be used at each level for consistency with the internal homotopy structure connecting the different levels.


\subsection{The homotopy lattice cutoff}\label{HLcutoffSubSect}

Now that HLGFs have been formally introduced, it is clear that $X$ a cellular decomposition of the base space induces a cutoff map sending any gauge field in the continuum $A \in {\cal A}^\infty_{M, G}$ to its corresponding HLGF $A^{HL}= c_X (A) \in {\cal A}^{HL}_{X_\ast , G}$ constructed evaluating the higher parallel transport of $A$ on higher dimensional globes fitting in $X_\ast$ the skeleton of the appropriate dimension
\[
c_X: {\cal A}^\infty_{M, G} \to {\cal A}^{HL}_{X_\ast , G} . 
\]
We will describe the space of HLGFs in detail in the second part of this series.

\subsection{A set of generators for HLGFs: ELGFs}\label{ELGFs}

In this section, except where we explicitly say otherwise, we assume that the cellular decomposition $X$ is a triangulation by a simplicial set. 

Extended Lattice Gauge Fields (ELGFs) were introduced by Claudio Meneses in collaboration with one of the authors of this paper \cite{Meneses:2017vqn, Meneses:2019bok}. 
The motivation and the main result were to refine the lattice cutoff for gauge fields in such a way that the resulting  extended lattice gauge fields retain the ability of characterizing a $G$-bundle (up to bundle equivalence) after the cutoff%
\footnote{Article \cite{Meneses:2017vqn} proposes two notions of extended lattice gauge fields. The fields in the first of them tautologically characterize a bundle, and the second turns out to characterize the first for two and three dimensional base spaces, but not in the general case. 
The second notion is the one further developed in \cite{Meneses:2019bok}, and it is equivalent to what we call ELGFs in this paper.}. 
%
%
In this subsection we describe how the version of ELGFs given in \cite{Meneses:2019bok} can be understood as HLGFs evaluated on 
a special set of elements of $\rho^{1,\bigcirc} X_\ast$, the set of simplices of paths. 
The idea was that, due to the special properties of the set of simplices of paths, evaluation of a homotopy aware discretized gauge field on simplices of paths completely determines the field, and that for each set of evaluations satisfying a set of compatibility conditions to be imposed in a particular order, 
the existence of the gauge field in the continuum whose discretization has the given set of evaluations 
can be ``guaranteed inductively'' in a sense that will become clear later on. 
Below we will describe the properties of the set of simplices of paths in the language of the globular algebra developed in previous sections. 
The property of the set of simplices of paths $\{ \Gamma_\sigma \}_{\sigma \in X_k | k\geq 1}$ is that it is a set of generators of $\rho^{1,\bigcirc} X_\ast$, which furthermore is free relative to their boundaries. 
That algebraic language was not used in \cite{Meneses:2017vqn, Meneses:2019bok}, but the inherited properties of the set of evaluations were always the motivation of the definition of the version of ELGFs given in \cite{Meneses:2019bok}. 

The original work on ELGFs did not provide a notion of higher homotopy parallel transport, nor the complete algebraic toolbox for gluing simplices of paths. The mentioned relationship will let us import the topological results of ELGFs to the algebraically superior structure of HLGFs.

{\em Simplices of paths}\\
Simplices of paths were described in Section \ref{PrelimExample}. The 0-simplex of paths $\Gamma_{ij}$ is a path (i.e. 1-globe) whose image is contained in the 1-simplex of $X$ with ordered vertex set $[ij]$. In our convention, it has $j$ as source and $i$ as target. 
The 1-simplex of paths $\Gamma_{ijk}$ is a 1-globe of paths whose image is contained in the 2-simplex of $X$ with ordered vertex set $[ijk]$. It represents a 1-homotopy of paths, all of them with $k$ as source and $j$ as target. The homotopy uses the geometric 1-simplex $|[ij]|$ as parameter space: Every $x \in |[ij]|$ determines a path passing by $x$ on its way from $k$ to $j$.
The globular structure of $\Gamma_{ijk}$ was explicitly given in Section \ref{Globes+ThinHomot} (see Figure \ref{Gamma{ijk}As2globes}). 
Similarly, the 2-simplex of paths $\Gamma_{ijkl}$ is a 2-globe of paths with image in the 3-simplex with ordered set of vertices $[ijkl]$, and it represents a homotopy of paths using the geometric 2-simplex $|[ijk]|$ as parameter space (which has the structure of a 2-globe). 

The general construction directly extends the one shown in Figure \ref{Gamma{ijk}As2globes}, making it the case $k=1$. 
Write $\Gamma^k=N^k\circ M^k$ where $M^k:G^{k+1}\to Y^{k+1}$ is a homeomorphism from the standard $(k+1)$-globe to a closed $(k+1)$-ball presented
as the union of two $(k+1)$-simplices with vertices $[1,\ldots,k+1,a]$ and $[1,\ldots,k+1,b]$. 
The choice of $M^k$ must be compatible with the choice of $M^{k-1}$. Our choice satisfies $M^k|_{e_k^-}=M^{k-1}$ 
(where we have identified $Y^k$ with the union of the two $k$-simplices in the boundary of $Y^{k+1}$ by sending $j$ to $j+1$ for $1\leq j\leq k$ and fixing $a$ and $b$), and $M^k$
sends the center $(1,0,\ldots,0)$ of $e_k^+$ to vertex $1$. 
On the other hand $N$ is a simplicial map, $N^k|_{[1,\ldots , k+1, a]}$ is the injective map 
$N^k(a) =i_{k+1}$, $N^k(j)= i_{j-1}$ for $1 \leq j \leq k+1$, and $N^k(b) = N^k(k+1) = i_k$.
If we want to include the name of the (k+1)-simplex $\sigma = [v_0, \ldots , v_{k+1}]$ we will write $\Gamma^k_\sigma$ or $\Gamma_\sigma$ if the dimension of the simplex doesn't need to be specified.

The original construction presented in \cite{Meneses:2019bok} uses a triangulation by a simplicial set constructed as the barycentric subdivision of a simplicial complex. 
In that reference a k-simplex of paths contained in simplex $\nu$ of $X$ is denoted by $\Gamma_{\tau \nu}$ where $\tau$ is the k-simplex serving as parameter space for the homotopy; the 2-simplex of paths $\Gamma_{ijkl}$ would have been written as $\Gamma_{\tau \nu}$ with $\nu = [ijkl]$ and $\tau = [ijk]$. 
In \cite{Orendain:2023tly} we use a similar notation (reversing the roles of the Greek letters), $\Gamma_{\nu\tau}$ with $\nu \subset \tau$ and the parameter space for the homotopy being $\nu$.

{\em The set of simplices of paths is generating and cell free relative to their boundaries}\\
Let us formally state the property of being a generating set that is cell free relative to the boundary. 

First notice that $\rho^{1,\bigcirc} X_\ast \supset \rho^{1,\bigcirc} (X_k)_\ast$ for $k \geq 1$, and $\rho^{1,\bigcirc} X_\ast$ can be constructed as the colimit of the nested infinity groupoids $\rho^{1,\bigcirc} (X_k)_\ast$ as $k$ goes to infinity. 

Now, the property of being generating and 
cell free relative to the boundary can be stated as follows: \\
A set $\{ p_\sigma \in \rho^{1,\bigcirc} X_\ast \}_{\sigma \in X_k | k\geq 1}$ is a set of free generators relative to their boundaries iff 
$\rho^{1,\bigcirc} (X_{k+1})_\ast$ can be obtained from 
$\rho^{1,\bigcirc} (X_k)_\ast \subset \rho^{1,\bigcirc} (X_{k+1})_\ast$ by adjoining one generator per (k+1) simplex and the added generator $p_\sigma$ is bound to comply with compatibility conditions only with elements of $\rho^{1,\bigcirc} (\partial \sigma)_\ast \subset  \rho^{1,\bigcirc} (X_k)_\ast$. 
Furthermore, 
it should happen that e
very morphism defined on the preceding skeleton extends uniquely after specifying the image
of each new generator, provided 
consistency with the definition on the lower dimensional skeleton.

It is known that $\rho^\bigcirc X_\ast$ has the property of cell freeness relative to the boundary, but for another set of generators associated to cells \cite{BrownNAT}. In fact, it is a key property behind the Higher Homotopy Seifert-van Kampen theorem which is the central theorem and main motivation of the strict infinity groupoid of Brown and collaborators. 

Our modified set focusing on globes of paths does not make any difference for this property; what we need to do is prove that our set $\{ \Gamma_\sigma \}_{\sigma \in X_k | k\geq 1}$ satisfies the property. 

Here is an algebraic proof. 
We only need to prove that any element $p_\sigma$ of the known set satisfying the property can be written in terms of $\Gamma_\sigma$ and elements of $\rho^{1,\bigcirc} (\partial \sigma)_\ast$. 

For 1-simplices, the sets of generators coincide. Now we consider $\sigma$ a (k+1)-simplex with $k\geq 1$. 

The globular faces $d_k^\pm\Gamma_\sigma$ belong to $\rho^{1,\bigcirc} (\partial \sigma)_\ast$. 

The globe of paths $q_\sigma = s_{k, k+1}(-_0d_k^-\Gamma_\sigma) \circ_0 \Gamma_\sigma$ has the properties $d_0^- q_\sigma = d_0^+ q_\sigma = d_0^- \Gamma_\sigma = v$, $d_k^- q_\sigma = s_{0,k} v$, which make it 
determine a class in $\pi_{k+1}(\sigma, \partial \sigma , v) \cong \Z$, realized as an abelian subgroup of $\rho^{1,\bigcirc} \sigma_\ast$ using $\circ_0$ as its product. 

Since the class of $q_\sigma$ has degree $\pm 1$, it generates the relative homotopy group. This is the key ingredient of our proof. 

Since $\sigma$ is a (k+1)-simplex with $k\geq 1$, its boundary $\partial \sigma$ is path connected, and we can construct $\tilde{p}_\sigma$ a whiskered version of $p_\sigma$ inducing a class in the same subgroup $\pi_{k+1}(\sigma, \partial \sigma , v) \subset \rho^{1,\bigcirc} \sigma_\ast$ 
\[
\tilde{p}_\sigma = s_{1, k+1}(-_0 c) \circ_0 \hat{p}_\sigma \circ_0  s_{1, k+1}(c) , 
\]
with $\hat{p}_\sigma = s_{k, k+1}(-_0 d_k^- p_\sigma) \circ_0 p_\sigma$ and $c$ a path in $\partial \sigma$ with $d_0^- c = v$ and $d_0^+ c = d_0^- p_\sigma$. 

Since $q_\sigma$ generates, there is an integer $m$ such that $\tilde{p}_\sigma = m q_\sigma$, in the sense of $\circ_0$ composing $q_\sigma$ with itself $m$ times. (In fact, $\tilde{p}_\sigma$ has degree $1$, which implies $m = \pm 1$.) Since $-_0$ is an inverse map for the $\circ_0$ composition, all the operations involved in the above definitions are explicitly invertible, and a short exercise in globular algebra explicitly writes $p_\sigma$ in terms of $\Gamma_\sigma$ and degeneracies over globes belonging to $\rho^{1,\bigcirc} (\partial \sigma)_\ast$, completing the proof.

{\em The evaluation of HLGFs on simplices of paths}\\
Let us define 
\[
A^{EL}_{i_0 i_1 \ldots i_{k+1}} = A^{HL}(\Gamma_{i_0 i_1 \ldots i_{k+1}}) \in (\rho^\bigcirc G)_k
\]
for every $0\leq k \leq n - 1= \dim X - 1$. 
This is our notation for the evaluation of $A^{HL}$ on all the simplices of paths in $X$. The set of evaluations is an ELGF. 
Because the set of simplices of paths is a generating set for globes in $X$ we know that an ELGF determines a HLGF. The property of being cell free relative to their boundaries will let us write an ELGF as a data set subject to a set of conditions, and in this way specify a unique HLGF.

{\em The ELGF as a generating consistent data set}\\
Data $\{ A^{EL}_{i_0 i_1} \}$ corresponding to path simplices fitting in $X_1$ is free. Once consistent data for path simplices fitting in $X_k$ (with $k\geq 1$) has been specified, candidate data $\{ A^{EL}_{i_0 i_1 \ldots i_{k+1}} \}$
corresponding to k-simplices of paths in $X_{k+1}$ is subject to the following conditions:%
\footnote{This is exactly the set of conditions written in references \cite{Meneses:2017vqn, Meneses:2019bok}.}
\begin{enumerate}
	\item Compatibility with lower dimensional data \\
    For every $\Gamma_{i_0 i_1 \ldots i_{k+1}}$ in $X_{k+1}$ 
\begin{eqnarray}\label{ELGFsFaceCompatibility}
		&d_{k-1}^- A^{EL}_{i_0 i_1 \ldots i_{k+1}} = A^{EL}(d_k^- \Gamma_{i_0 i_1 \ldots i_{k+1}}) , \quad \nonumber \\
		&d_{k-1}^+ A^{EL}_{i_0 i_1 \ldots i_{k+1}} = A^{EL}(d_k^+ \Gamma_{i_0 i_1 \ldots i_{k+1}}) ,  
	\end{eqnarray}
    where the right hand side is read as a word on the specified data set at $X_k$. 
	\item Extendibility \\
    If the dimension of the base is larger than $k+1$, for every $\Gamma_{i_0 i_1 \ldots i_{k+2}}$ in $X_{k+2}$ 
    \begin{equation} \label{ELGFsInteriorExtendibility}
		A^{EL}(d_{k+1}^- \Gamma_{i_0 i_1 \ldots i_{k+2}}) = A^{EL}(d_{k+1}^+ \Gamma_{i_0 i_1 \ldots i_{k+2}}) ,  
	\end{equation}
    where both sides of the equations are read as words on the candidate data set. The condition says that the two distinct k-globes of paths in the argument (interpolating between the same pair of source and target (k-1)-globes of paths) leading to two separate strings of words in the data set which coincide after each string is properly assembled in $\rho^\bigcirc G$, as it would happen when evaluating homotopies of parallel transport maps of a gauge field in a shell with a filling. 

    Notice that when candidate data corresponding to $X_{k+2}$ are considered (at the next step in the induction process), after they satisfy condition \eqref{ELGFsFaceCompatibility} condition \eqref{ELGFsInteriorExtendibility} becomes an algebraic identity. 
\end{enumerate}

We need to address now the issue of completeness of the above inductive set of conditions; that the data set is subject to those conditions and no extra requirements in order to be consistent as a set of evaluations of a strict infinity groupoid morphism. Completeness follows from the property of cell freeness relative to the boundary, which translates to precisely the statement that apart from conditions \eqref{ELGFsFaceCompatibility} no more conditions are required to extend the morphism from 
$\rho^{1,\bigcirc} (X_k)_\ast$ to $\rho^{1,\bigcirc} (X_{k+1})_\ast$. The role played by conditions \eqref{ELGFsInteriorExtendibility} is to guarantee that, if needed, the data set can be extended to the next skeleton.

Codimension-1 source and target of a simplex of paths, play an important role on the definition of ELGFs as a consistent data set, in particular in condition \eqref{ELGFsFaceCompatibility}. 
An explicit expression would be essential for any computer implementation. There are many equivalent ways to write them. Here we introduce a bit of extra notation to give a compact version. 

The idea is simple, 
our definition of $\Gamma_{i_0\ldots i_{k+1}}$ with its globular structure has $d_k^-\Gamma_{i_0\ldots i_{k+1}}$ as the single simplex of paths $\Gamma_{i_1\ldots i_{k+1}}$, while $d_k^+\Gamma_{i_0\ldots i_{k+1}}$ is a globe of paths constructed appropriately gluing the simplices of paths associated to all its codimension-1 faces that touch vertex $i_0$. 
That gluing would be simple to do if the pieces associated to the codimension-1 faces containing $i_0$ had completely degenerate codimension-1 source over $i_0$. Since whiskering the original pieces can bring them to have this property, and it is an invertible map, we will use it as an intermediate step before gluing the resulting $i_0$-based globes using the $\circ_0$ operation. Since this is the context, in the following formulas the product uses $\circ_0$ (writing factors left to right as $j$ increases) and the exponent $-1$ denotes the inverse $-_0$. 

Let $I=(i_0,\ldots,i_{k+1})$ be ordered increasingly, and let
$I_j$ be the list obtained by omitting $i_j$. 
Simplices of paths containing $i_0$ are $\Gamma_{I_j}$ labeled by a value of $j$. 
Then
\[
d_k^-\Gamma_I=\Gamma_{I_0}
=\Gamma_{i_1\ldots i_{k+1}}
\qquad (k\geq1).
\]
For $k=1$, the target is
\[
d_1^+\Gamma_{i_0i_1i_2}
=(-_0\Gamma_{i_0i_1})\circ_0\Gamma_{i_0i_2}.
\]

For $k\geq2$, let $J_j$ be $I_j$ with its first entry omitted,
for $0\leq j\leq k+1$. 
We first modify each face containing
$i_0$ to have its $(k-1)$-source be completely degenerate and located at the source vertex of $\Gamma_{I_j}$; this vertex is $i_{k+1}$ for $1\leq j\leq k$, and $i_k$ for $j=k+1$. The result is 
\[
Q_j
=s_{k-1,k}(-_0\Gamma_{J_j})\circ_0\Gamma_{I_j},
\qquad 1\leq j\leq k+1.
\]

To transport all these globes to $i_0$, put
\[
c=\Gamma_{i_0i_{k+1}},
\qquad
h=\Gamma_{i_ki_{k+1}},
\qquad
c_j=
\begin{cases}
c, & 1\leq j\leq k,\\[2pt]
c\circ_0(-_0h), & j=k+1.
\end{cases}
\]
Each $c_j$ runs from the source vertex of the corresponding
face to $i_0$. Define
\[
\widehat Q_j
=s_{1,k}(c_j)\circ_0 Q_j\circ_0 s_{1,k}(-_0c_j).
\]
These globes satisfy
\[
d_{k-1}^-\widehat Q_j=s_{0,k-1}(i_0),
\qquad
d_0^-\widehat Q_j=d_0^+\widehat Q_j=i_0,
\]
so they and their $\circ_0$ inverses can be composed at $i_0$.

The desired expression for $d_k^+\Gamma_I$ is obtained by transporting the result back to $i_{k+1}$, and restoring 
its $(k-1)$-source before our modifications:
\begin{align*}
d_k^+\Gamma_I
={}&s_{k-1,k}(\Gamma_{J_0})
    \circ_0 s_{1,k}(-_0c)\\
 &{}\circ_0
    \left(
      \widehat Q_{k+1}^{\,(-1)^k}
      \circ_0
      \prod_{j=1}^{k}
      \widehat Q_j^{\,(-1)^{j+1}}
    \right)
    \circ_0 s_{1,k}(c).
\end{align*}
The factor inside the parenthesis has completely degenerate $(k-1)$-source
over $i_0$. The outer factors restore its boundary to 
\[
d_{k-1}^-d_k^+\Gamma_I=\Gamma_{J_0},
\qquad
d_0^-d_k^+\Gamma_I=i_{k+1},
\qquad
d_0^+d_k^+\Gamma_I=i_k.
\]

On the tetrahedron with vertices $[0,1,2,3]$, after 
cancellation the general formula yields 
\begin{align*}
d_2^-\Gamma_{0123}
 &=\Gamma_{123},\\
d_2^+\Gamma_{0123}
 &=s_{1,2}(-_0\Gamma_{12})
   \circ_0\Gamma_{012}
   \circ_0\Gamma_{023}\\
 &\qquad{}\circ_0(-_0\Gamma_{013})
   \circ_0s_{1,2}(\Gamma_{13}).
\end{align*}

Another consequence of the above construction of ELGFs as sets of consistent data with the organization described above is a proof that, for base spaces of dimension up to 4, for every HLGF there are gauge fields in the continuum realizing them. Every $A^{HL}$ induces one $A^{EL}$, and $A^{EL}$ is a local data set satisfying conditions \eqref{ELGFsFaceCompatibility}, \eqref{ELGFsInteriorExtendibility}. 
For a given $A^{HL}$, the question of existence of $A$ gauge field in the continuum with $A^{HL}(A)= A^{HL}$ can be addressed at stages labeled by the skeleta: Is there an $A$ on the restricted base $|X_k|$ such that it is compatible with $A^{HL}|_{|X_k|}$? If the answer is affirmative, the same question for the skeleton of one dimension higher can be addressed as an extendibility question. 
That extendibility question at each cell 
is the question of existence of a homotopy, which 
for bases up to dimension 4 is precisely \eqref{ELGFsInteriorExtendibility}. 
Then having $A^{EL}(A^{HL})$ a data set satisfying the condition at each (k+1)-cell given the set of boundary conditions imposed by the known field at the boundary of the cell guarantees extendibility at each cell. 
Cell freeness relative to the boundary tells us that the set of conditions is sufficient, completing the proof. 

For 5-dimensional bases the obstructions to cell extendibility 
can be deformed by whiskering to 
live in $\pi_3 G$, which is not detected by our $\rho^\bigcirc G$. For 4-dimensional bases the obstructions live on $\pi_2 G$ which is trivial for any Lie group. In the 4-dimensional case, there are fields in the continuum $A$ realizing $A^{HL}$, but the HLGF does not determine the bundle class for every structure group. See Section \ref{TopChargeSection}.

\subsection{General higher gauge fields on the homotopy lattice}\label{GeneralHigherGaugeFields}

In this article, our main objective is to describe ordinary gauge fields within a cutoff as explained previously. In this subsection we deviate from this goal to consider ``higher gauge fields'' not necessarily associated with relative higher homotopy classes of ordinary parallel transport. 

Considering our definition \eqref{PT} of a HLGF as a section of a groupoid with higher homotopy internal structure, 
\[
PT: \rho^{1,\bigcirc} X_\ast  \to At^{1,\bigcirc}(X_\ast; \rho^\bigcirc G), 
\]
we see that a route for generalizing our higher gauge fields is to find infinity groupoids $F$, $K$ replacing $\rho^\bigcirc {\cal F}$ and $\rho^\bigcirc G$ respectively, and with the additional internal group structure in $K$ and the transitive and free right action of $K$ on $F$. 

From the construction given here, we see that the resulting higher gauge fields would come with a notion of higher parallel transport along globes of paths: The $\circ_0$ gluing map of $\rho^\bigcirc X_\ast$ would be in correspondence with the internal multiplication in the replacement of $\rho^\bigcirc G$. 

This generalization exercise was studied, using definition \eqref{Aphi}, in \cite{Orendain:2023tly}; we found agreement with an existing notion of higher gauge fields (using 2-groups) in the lattice \cite{Pfeiffer:2003je}. Additionally, we found in \cite{Orendain:2023tly} that any such generalization $(F, K)$ can be realized by an appropriate filtered group $G$. That is, given $(F, K)$ with the specified structure, there is a filtered group $G$ such that $K \cong \rho^\bigcirc G$ and $F \cong \rho^\bigcirc TG$ where $TG$ is a torsor for $G$.

\section{The topological charge and other topological considerations}\label{TopChargeSection}

The topological charge is a functional of the gauge field that is a topological invariant of the $G$-bundle. Since it happens that the evaluations of the functional can only achieve values in a discrete set, the functional may carry non trivial information only if the space of gauge fields has more than one connected component; 
since the space of gauge fields over a given bundle is a connected space, the topological charge is relevant only if there is more than one $G$-bundle class over the given base space. 
Consider a physical situation of interest taking place in the base space $\R^n$, representing spacetime, accompanied by ``fall-off conditions'' for the gauge field ensuring at infinity the field is trivial up to gauge. In Euclidean quantum field theory the fall-off conditions are naturally implemented considering a compactification of the base space to $S^n$. 
In the continuum, the topological charge can be evaluated in several ways. The most common one is as an integral of a polynomial function of the curvature; we could try to regularize the curvature and the mentioned polynomial function in terms of HLGFs. The alternative way, which we review below, is to calculate the topological charge in $S^{2n}$ as the winding number of a map from $S^{2n-1}$ to $G$ representing the transition function between the local trivializations (constructed using the gauge field) over the two hemispheres. The equivalence of these two ways to calculate the topological charge is proven in \cite{kobayashi1996foundations}. 
At the end of Section \ref{Globes+ThinHomot} we saw that the homotopy lattice cutoff leading to HLGFs keeps homotopical information of the gauge field for base spaces of dimension 2 and 3. This will let us give a simple formula for the topological charge calculated as a 
class in $\pi_1 G$ for bases of dimension 2.

\subsection{The topological charge of gauge fields over $S^2$}\label{TopChargeS2}

We start with the example of $U(1) \simeq SO(2)$ HLGFs on $S^2$ with the triangulation $X$ described in Section \ref{PrelimExample}. 

The main idea of the calculation of the topological charge as a winding number in the continuum \cite{kobayashi1996foundations} has the following ingredients: \\
(i) Separate the base $S^2$ into two charts $H_S$, $H_N$ with the topology of the disc intersecting in a neighborhood with the topology of $S^1_{\rm Equator} \times (-\epsilon, \epsilon)$. \\
(ii) Use the gauge field and a path system to set up a trivialization over $H_S$. The path system is the assignment of a path $m_S^x$ to every point $x \in H_S$ from the base point $S$ (the South Pole) to $x$. Then the gauge field $A$ parallel transports any $u_S\in {\cal F}_S \simeq U(1)$ to $A (m_S^x) u_S \in {\cal F}_x$. Moreover, since the parallel transport map $A (m_S^x): {\cal F}_S \to {\cal F}_x$ is 1 to 1, any point in ${\cal F}_x$ can be parametrized using ${\cal F}_S$. This procedure determines a trivialization over $H_S$; the analogous procedure yields a trivialization over $H_N$. \\
(iii) The transition function $\psi_{SN}: H_S \cap H_N \to U(1)\simeq S^1$ restricted to ${\rm Eq}= S^1_{\rm Equator} \times \{ 0 \}$ is a continuous map determined by the gauge field $A$ characterizing the bundle 
\[
\psi_{SN}|_{\rm Eq}(A)(x) = \left( A (m_N^x)\right)^{-1} \cdot A (m_S^x), 
\]
where in this formula we see each factor in the righthand side as valued in $G$. 
A crucial property is that a homotopy between two such transition functions $\psi_{SN}|_{\rm Eq}$ determines a bundle map between the corresponding bundles, making them equivalent. Homotopy classes are classified by their winding number, and the topological charge is defined to be the winding number 
\begin{equation}\label{TopCharge2dContinuum}
	Q(A) = W (\psi_{SN}|_{\rm Eq}(A)) . 
\end{equation}

For the HLGF the important principle is that the homotopy lattice cutoff keeps the exact homotopical information when it is cast in a format that is compatible with their algebraic structure. 
Thus, the regularization of \eqref{TopCharge2dContinuum} for HLGFs follows four steps: 
\begin{enumerate}
	\item 
	Choose a triangulation $X$ with one vertex in $S$, another vertex in $N$ and such that the Equator is contained in $X_1$. 
	\item
	Realize $\psi_{SN}|_{\rm Eq}$ as the evaluation of $A$ the gauge field on $m$ the singular 1-globe of paths (the family of meridians) with source in $S$, target in $N$ and parametrized by a singular 1-globe winding once around the Equator 
	\[
	\psi_{SN}|_{\rm Eq} = A(m) . 
	\]
	\item
	Then the evaluation of the 
	HLGF $A^{HL} = c_X (A)$ on $[m]$ the 1-globe of paths in $X$ (up to thin homotopy) is precisely the relative homotopy class of $\psi_{SN}|_{\rm Eq}$ 
	\[
	[\psi_{SN}|_{\rm Eq}] = [A(m)]=(c_X (A)) ([m]). 
	\]
	\item
	Since $d_1^- [m] = d_1^+ [m]$ the evaluation $(c_X (A)) ([m])$ is a 1-globe in $G=U(1)$ with its source equal to its target, and its homotopy class relative to endpoints is characterized by its winding number 
	\[
	W (A^{HL} ([m])) = W ((c_X (A)) ([m])) = W (\psi_{SN}|_{\rm Eq}(A)) = Q(A) . 
	\]
\end{enumerate}

This result captures the essence of what HLGFs can do that standard lattice gauge fields can't. 

Below we present a more abstract version of the same idea, which shows the power and versatility of the algebraic machinery presented in this article: \\
A simple observation lets us change the perspective of the calculation to base it on higher parallel transport. $A^{HL}$ determines parallel transport along 1-globes of paths transporting 1-globes of initial conditions on the 0-source of the globe of paths. Thus, a constant initial condition $s_{0,1} u \in \rho^\bigcirc_1 {\cal F}_S$ is transported along $[m]$ from the South Pole to the North Pole. The result is a 1-globe $A^{HL} ([m])(s_{0,1} u) \in \rho^\bigcirc_1 {\cal F}_N$ with equal source and target; its homotopy class is determined by its winding number, and this calculation coincides with the topological charge 
\[
	W (A^{HL} ([m])(s_{0,1} u)) = Q(A^{HL})  
\]
in the sense that if $A^{HL}= c_X (A)$ then $Q(A^{HL}) = Q(A)$. 

Here is yet another view on the calculation of the topological charge using HLGFs. 
Notice that the 1-globe of paths $m$ used above is constructed from a circle $E$ embedded in $X_1$ and two points lying on opposite sides of the circle (playing the role of the Poles). That circle can be given a base point $x$ to host a representation of $\pi_1(S^1, x) \subset \rho^\bigcirc X_\ast$. The base point determines $Gr$, a ``Greenwich Meridian'' also embedded in $X_1$. 
Given any element $n \in \Z \simeq \pi_1(E, x)$, we can choose a based curve $c_n$ inside the circle representing class $n$, and the two ``Poles'' to parametrize a 1-globe of paths, the resulting 1-globe of paths will cover $S^2$ $n$ times. Evaluation of $A^{HL}$ on the constructed 1-globes of paths induces a group morphism 
\begin{equation}\label{TopCharge2dHLGFs}
	Q(A^{HL}) : \pi_1(E, x) \to \pi_1 (U(1), A^{HL}(Gr)) \simeq \Z 
\end{equation}
which in this case is characterized by the multiplication by an integer. For other gauge groups, like $G=SO(3)$, this definition has a clear extension. 

Notice that our definition of $Q(A^{HL})$ implies a collection of equivalent formulas to evaluate the topological charge in terms of local data. An ELGF $A^{EL}$ is local data characterizing a HLGF $A^{HL}(A^{EL})$, and the globular algebra gives us precise formulas to evaluate $Q(A^{EL})$. 
In Section \ref{PrelimExample} we worked out the example of $A^{HL}_{LC}$ the $SO(2) \simeq U(1)$ gauge field induced by the Levi-Civita connection on $S^2$ restricted to parallel transport unit vectors. We found that $QA^{HL}_{LC}= 2$. 

It is easy to show that for any given $n \in \pi_1 SO(2) = \Z$ there are 
gauge fields over $S^2$ with $Q(A) = n$. Then it is clear that also there are HLGFs over a trinagulated $S^2$ with $Q(A^{HL}) = n$.


Let us briefly explore what happens if we try to calculate the topological charge for a gauge field in the continuum $A$ on the 4-dimensional base $S^4$ as a winding number of a transition function $\psi_{SN}|_{\rm Eq}(A): S^3 \to G$. The result is that the winding number is not captured by $\rho^\bigcirc G$ for a filtered space $G$ with the trivial filtration, as shown in the argument at the end of Section \ref{Globes+ThinHomot}. Thus, $A^{HL}(A)$ does not retain that information present in the continuum gauge field $A$.

We should also mention that Chen has announced a program for calculating the topological charge in four dimensions \cite{Chen:2024ddr, Zhang:2024sgm}. In Section \ref{Summary+Discussion} we discuss the relation between his approach and ours.

\subsection{The bundle induced by a HLGF over 2 or 3 dimensional base spaces}\label{BundleOver2Dor3D}

\begin{figure}[ht] \centering
    \includegraphics{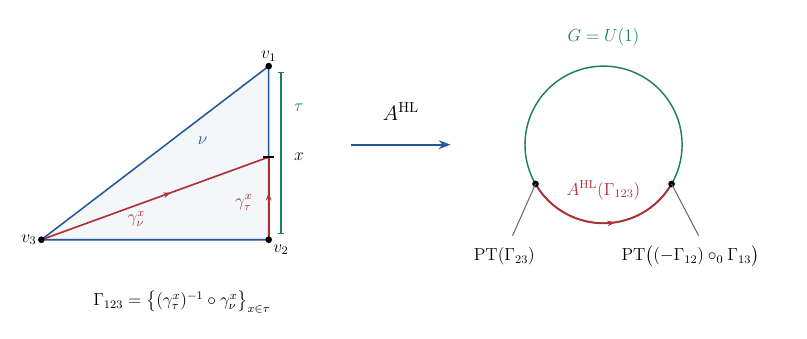}
    \caption{}
    \label{FigX}
\end{figure}

{\em The bundle induced by a HLGF} \\
We saw that a HLGF $A^{HL}$ determines an extended lattice gauge field $A^{EL}(A^{HL})$ simply by evaluating on simplices of paths. 
On the other hand, on a triangulated base space of dimension 2 or 3 
an extended lattice gauge field with gauge group $G$ induces a $G$-bundle $\pi_{A^{EL}}$ \cite{Meneses:2019bok, Meneses:2017vqn} 
. Therefore, for base spaces of dimension 2 or 3, a HLGF $A^{HL}$ induces a $G$-bundle 
\[
A^{HL} \longmapsto \pi_{A^{HL}} . 
\]
The proof given in \cite{Meneses:2019bok} explicitly constructs local trivializations and transition functions. The transition functions are determined by the parallel transport map induced by the gauge field in the continuum; then, as shown in Section \ref{Globes+ThinHomot}, for base spaces of dimension 2 or 3 the HLGF captures the information of the parallel transport up to homotopy rel. vertices. 
Here we sketch its basic idea: \\
Consider $M$ triangulated by a simplicial set $X$. 
The first step states that, as proven at the end of Section \ref{ELGFs}, any $A^{EL}$ can be realized from the evaluation of a gauge field in the continuum $A$. 
This step does not tell us whether the continuum fields realizing a given $A^{EL}$ all determine the same bundle. 
In the second step, notice that each simplex $\nu$ of the triangulation may be considered as a chart, and with the aid of $A$ a gauge field in the continuum, a local trivialization is constructed: A path system on $\nu$ connects the maximal vertex of $\nu$ to each point $x\in \nu$ by a path $\gamma_\nu^x$. Then, once an initial condition $u_0$ in the fiber over the maximal vertex is chosen, $A(\gamma_\nu^x)$ transports $u_0$ to a fiber over $x$ defining the local trivialization. 
For points that lie in the boundary of a simplex $x \in \tau \subset \nu$ (with $\tau$ not containing the maximal vertex of $\nu$) the parallel transport maps $A(\gamma_\nu^x), A(\gamma_{\tau}^x)$ tell us how to glue the two local trivializations. In that way, a bundle over $M$ is determined by gluing the trivial bundles over the simplices of $X$. 
The standard lattice cutoff tells us how to glue the mentioned trivial bundles over simplices at the vertices only. The homotopy lattice cutoff refines the standard lattice cutoff in that (for base spaces of dimension 2 or 3) $A^{HL}(A)= c_X(A)$ keeps the homotopy class of the gluing maps relative to the evaluation at vertices 
(see Figure \ref{FigX}, illustrating a case with a two-dimensional base and $G=U(1)$). 
The stored homotopy information is enough to determine the class of the bundle because two such continuum gauge fields $A, A'$ which may be different but induce the same $A^{EL}$ would induce gluing maps that are homotopic relative to vertices, and those homotopies relative to vertices can be used to construct bundle equivalence maps. 

As shown in the argument at the end of Section \ref{Globes+ThinHomot}, 
the topological charge associated to $\pi_3 G$ for gauge fields on a 4-dimensional base cannot be calculated using HLGFs, which also means that in those situations the bundle is not determined by a HLGF.

\section{Summary and discussion}\label{Summary+Discussion}

In this article we introduced HLGFs providing a geometrical motivation and a precise algebraic definition. We also pedagogically presented the necessary algebraic framework backing up our proposal. 
We showed that, for base spaces of dimension 2 or 3, a HLGF determines a principal bundle. 
We also gave exact explicit formulas for the topological charge of a HLGF in dimension 2.

In a base space triangulated by a simplicial set, 
HLGFs evaluated on simplices of paths determine a set of generators for HLGFs. 
In previous work of one of the authors in collaboration with Claudio Meneses, these generators were called extended lattice gauge fields \cite{Meneses:2017vqn, Meneses:2019bok}. 
This work is the direct precursor of what we presented here. 
We saw that in dimension 2 HLGFs reproduce a previous proposal given by Pfeiffer \cite{Pfeiffer:2003je}. 

Let us briefly mention the reason making our claim of characterizing a principal bundle over the base manifold be valid only for bases of dimension 2 or 3. 
The obstacle for a stronger result comes from our use of a trivial filtration in the gauge group $G$. The mathematical framework mixing homotopy with the group structure in $G$ requires that the first level of the filtration $G_0$ be a subgroup of $G$. Mathematically speaking we could work with a discrete subgroup $G_0 \subset G$, but physics prefers the choice $G_0 = G$, and we stick with it. 
Questions regarding $\pi_k Y$ involve maps $S^k \to Y$. At the algebraic level, the framework that we used to talk about higher homotopy is based on maps $G^k \to Y$ whose domain is the standard k-globe, which has the topology of a disk, instead of the sphere. Since the maps are considered up to homotopy relative to vertices in the case $k=1$ when the image of the two vertices of $G^1$ coincide, we obtain the information about the fundamental group. For higher dimensional globes, if the whole boundary of $G^k$ were fixed during the homotopy, we could capture the information about $\pi_k Y$. This happens when we restrict to filtered maps $G^k \to Y$ and the filtration in $Y$ 
does not allow homotopies unwinding the image of $\partial G^k$. The curious reader could work out the instructive example of studying homotopy classes of filtered maps $G^2 \to S^2$ relative to vertices with respect to the trivial filtration of $S^2$; those homotopy classes of maps cannot see the second homotopy group of the sphere. This is what fails due to our use of a trivial filtration in $G$. Then we only see $\pi_k G$ for $k=1$. Fortunately for any Lie group $\pi_2 G$ is trivial; thus, our failure starts at the detection of $\pi_3 G$, which means that for bases of dimension 3 or 2 we see all relevant homotopical aspects of the gauge field. 

A detailed explanation will be included in Ivan Sanchez' PhD thesis (in preparation). 

For purely mathematical interests, we could use a variant of our HLGFs using a nontrivial filtration in the gauge group to calculate characteristic classes. 

At the end of Section 4.1 we mentioned the recently announced  program of Chen \cite{Chen:2024ddr, Zhang:2024sgm} to develop a higher category theory framework for lattice gauge fields to be able to calculate the topological charge. Clearly, that program and our work share motivation. They also share the use of higher category theory. Our notion of higher homotopy parallel transport, and its properties, cannot be understood outside of its higher dimensional groupoid context. On the other hand, Chen’s program involves elements that are outside of the mathematical scope of our work. 
The math structure on which HLGFs are based is already fully developed. It is a rich powerful framework, that lets us do certain things very efficiently and elegantly. 
Certain limitations arise because physics calls for certain structures, like having a trivial filtration in the gauge group $G$, and the math tells us what we can do with those restrictions. 
It is fortunate that in dimension 4, the dimension of spacetime, boundary data lives in dimension 3, where the limitations of our framework do not show. Thus, in a TQFT-like framework, where codimension-1 manifolds are assigned spaces of states constructed from boundary data, our framework may give us everything we need to construct those boundary state spaces. This is already very promising because many structural issues depend mostly on the space of states, and on observables depending purely on boundary data. 
On the other hand, if in the amplitude of a particular theory the bundle structure plays an important role, or if we are studying a topological observable like the topological susceptibility, then we would need to supplement our framework with extra elements. 
Here is where the limitations of our framework reside. 
We have plans for addressing those limitations, but it could be that another framework, like the one that Chen is constructing, could provide an appropriate complement to our spaces of states. 

For most applications to 2 and 3 dimensional ordinary gauge field theory in condensed matter physics, our framework would be completely self-sufficient.

In the paper describing the second part of this work we will give structure to the space of fields that lets us define 
measures and amplitudes. 
We apply the ideas behind the mentioned results solving the local to global problem in higher dimensional homotopy \cite{BrownNAT} to construct the space of fields (and the Hilbert space associated to such spaces) from the space of fields associated to elementary blocks. 
We also use general theorems in higher algebra to justify our definition of the spaces corresponding to the continuum limit as the appropriate inverse limits. 
All these results were announced in the conference 17th Marcel Grossmann Meeting, and the proceedings paper \cite{Orendain:2024wfh} has been accepted for publication. 
We also 
show that our framework refines standard lattice gauge theory in the sense that it agrees in its predictions regarding observables following from their 1-dimensional parallel transport, but we have access to a wider algebra of observables following from our higher dimensional parallel transport. In the case of 2d Yang-Mills theory, for example, we have a lattice version based on the heat kernel measure that is quantum perfect. Since we can also calculate the topological charge exactly, we can calculate the topological susceptibility exactly \cite{Orendain:2024wfh}.

\appendix
\renewcommand{\theequation}{A\arabic{equation}}
\setcounter{equation}{0} 
\section{Appendix}\label{Appendix}

Here we write the structural relations of $\omega$-groupoids. The original reference is \cite{Algebraofcubes}, but the notation below follows a variant of \cite{brown2007new}, which is compatible with their notation for $\infty$-groupoids that we use here. The main distinction of the notation used below is that we compose from right to left. Two other changes are (i) that our symbols for the higher groupoid operations use the convention that we follow in the main body of the text for cubes, which we have chosen to be different from the one we use for globes, (ii) that we have added an extra subindex to the degeneracy operator because that ``extended notation'' makes our formula for gauge transformations more explicit.

$X$ is a cellular decomposition of $M$, and $X_*$
is the filtered topological space determined by its skeleta. Throughout,
$\alpha,\beta\in\{-,+\}$, and $-\alpha$ denotes the opposite sign.
All relations are understood for indices for which the displayed maps
are defined.

The globular face and degeneracy maps are
\begin{equation*}
\begin{array}{ll}
d_j^\pm:\rho^\bigcirc_nX_*
    \longrightarrow\rho^\bigcirc_jX_*,
    & 0\leq j<n,\\[4pt]
s_{i,n}:\rho^\bigcirc_iX_*
    \longrightarrow\rho^\bigcirc_nX_*,
    & 0\leq i\leq n,
\end{array}
\end{equation*}
where $s_{i,i}=Id_i$. They satisfy
\begin{equation*}
d_i^\alpha d_j^\beta=d_i^\alpha
\qquad\text{for }0\leq i<j<n,
\end{equation*}
and, for $0\leq i\leq n$ and $0\leq j<n$,
\begin{equation*}
d_j^\pm s_{i,n}:\rho^\bigcirc_iX_*
    \longrightarrow\rho^\bigcirc_jX_*,
\qquad
d_j^\pm s_{i,n}=
\left\{
\begin{array}{lcc}
d_j^\pm & \text{if} & j<i,\\[4pt]
Id_i    & \text{if} & j=i,\\[4pt]
s_{i,j} & \text{if} & j>i.
\end{array}
\right.
\end{equation*}
For $0\leq i\leq j\leq n$,
\begin{equation*}
s_{j,n}s_{i,j}=s_{i,n}.
\end{equation*}

For the cubical relations, write
$K_n=\rho^\square_nX_*$. The structure maps are
\begin{equation*}
\begin{array}{ll}
\partial_i^\alpha:K_n\longrightarrow K_{n-1},
    & 1\leq i\leq n,\\[4pt]
\varepsilon_i:K_{n-1}\longrightarrow K_n,
    & 1\leq i\leq n,\\[4pt]
\Gamma_i^\alpha:K_{n-1}\longrightarrow K_n,
    & 1\leq i<n.
\end{array}
\end{equation*}
We use Brown's sign convention for the two connections. On a
representative cube $c:I^{n-1}\longrightarrow X$, with
$I=[-1,1]$, they are induced by
\begin{equation*}
(\Gamma_i^\alpha c)(t_1,\ldots,t_n)
=
c(t_1,\ldots,t_{i-1},
  \mu^\alpha(t_i,t_{i+1}),t_{i+2},\ldots,t_n),
\end{equation*}
where
\begin{equation*}
\mu^-(s,t)=\max(s,t),
\qquad
\mu^+(s,t)=\min(s,t).
\end{equation*}
Thus the unsigned connection $\Gamma_i$ used in the main text is
$\Gamma_i^-$.

\begin{equation}
\begin{array}{ll}
\partial_i^\alpha\partial_j^\beta
    =\partial_{j-1}^\beta\partial_i^\alpha
    & i<j,\\[4pt]
\varepsilon_i\varepsilon_j
    =\varepsilon_{j+1}\varepsilon_i
    & i\leq j.
\end{array}
\end{equation}

\begin{equation}
\partial_i^\alpha\varepsilon_j=
\left\{
\begin{array}{lcc}
\varepsilon_{j-1}\partial_i^\alpha
    & \text{if} & i<j,\\[4pt]
Id  & \text{if} & i=j,\\[4pt]
\varepsilon_j\partial_{i-1}^\alpha
    & \text{if} & i>j.
\end{array}
\right.
\end{equation}

\begin{equation}
\begin{array}{ll}
\Gamma_i^\alpha\Gamma_j^\beta
    =\Gamma_{j+1}^\beta\Gamma_i^\alpha
    & i<j,\\[4pt]
\Gamma_i^\alpha\Gamma_i^\alpha
    =\Gamma_{i+1}^\alpha\Gamma_i^\alpha.
\end{array}
\end{equation}

\begin{equation}
\Gamma_i^\alpha\varepsilon_j=
\left\{
\begin{array}{lcc}
\varepsilon_{j+1}\Gamma_i^\alpha
    & \text{if} & i<j,\\[4pt]
\varepsilon_j\Gamma_{i-1}^\alpha
    & \text{if} & i>j.
\end{array}
\right.
\end{equation}

\begin{equation}
\Gamma_j^\alpha\varepsilon_j
    =\varepsilon_j^2
    =\varepsilon_{j+1}\varepsilon_j.
\end{equation}

\begin{equation}
\partial_i^\alpha\Gamma_j^\beta=
\left\{
\begin{array}{lcc}
\Gamma_{j-1}^\beta\partial_i^\alpha
    & \text{if} & i<j,\\[4pt]
\Gamma_j^\beta\partial_{i-1}^\alpha
    & \text{if} & i>j+1.
\end{array}
\right.
\end{equation}

\begin{equation}
\begin{array}{c}
\partial_j^\alpha\Gamma_j^\alpha
    =\partial_{j+1}^\alpha\Gamma_j^\alpha
    =Id,\\[4pt]
\partial_j^\alpha\Gamma_j^{-\alpha}
    =\partial_{j+1}^\alpha\Gamma_j^{-\alpha}
    =\varepsilon_j\partial_j^\alpha.
\end{array}
\end{equation}

If $a,b\in K_n$, then $a+_j b$ is defined if and only if
$\partial_j^-a=\partial_j^+b$: the cube $b$ is followed by the cube $a$.
Then
\begin{equation}
\left\{
\begin{array}{l}
\partial_j^-(a+_j b)=\partial_j^-b,\\[4pt]
\partial_j^+(a+_j b)=\partial_j^+a.
\end{array}
\right.
\end{equation}

Each direction $j$ gives a groupoid structure, with inverse $-_j a$.
In particular, whenever the composites are defined,
\begin{equation*}
\begin{array}{c}
(a+_j b)+_j c=a+_j(b+_j c),\\[4pt]
\varepsilon_j\partial_j^+a+_j a
    =a=a+_j\varepsilon_j\partial_j^-a,\\[4pt]
(-_j a)+_j a=\varepsilon_j\partial_j^-a,
\qquad
a+_j(-_j a)=\varepsilon_j\partial_j^+a.
\end{array}
\end{equation*}

\begin{equation}
\partial_i^\alpha(a+_j b)=
\left\{
\begin{array}{lcc}
\partial_i^\alpha a+_{j-1}\partial_i^\alpha b
    & \text{if} & i<j,\\[4pt]
\partial_i^\alpha a+_j\partial_i^\alpha b
    & \text{if} & i>j.
\end{array}
\right.
\end{equation}

The \textit{interchange laws}. If $i\neq j$, then
\begin{equation}
(a+_i b)+_j(c+_i d)
    =(a+_j c)+_i(b+_j d),
\end{equation}
whenever the indicated composites are defined.

In the matrix notation below, the horizontal direction points to the
left and the vertical direction points downward, as in the main text.
Thus the following array denotes
$(c+_i d)+_j(a+_i b)$, or equivalently
$(c+_j a)+_i(d+_j b)$:
\begin{equation*}
\left[
\begin{array}{lr}
a & b\\
c & d
\end{array}
\right]
\qquad
\xymatrix @-1pc @R=0.3cm @C=0.3cm {
i & *{} \ar[l] \ar[d]\\
*{} & j
}
\end{equation*}

\begin{equation}
\varepsilon_i(a+_j b)=
\left\{
\begin{array}{lcc}
\varepsilon_i a+_{j+1}\varepsilon_i b
    & \text{if} & i\leq j,\\[4pt]
\varepsilon_i a+_j\varepsilon_i b
    & \text{if} & i>j.
\end{array}
\right.
\end{equation}

\begin{equation}
\Gamma_i^\alpha(a+_j b)=
\left\{
\begin{array}{lcc}
\Gamma_i^\alpha a+_{j+1}\Gamma_i^\alpha b
    & \text{if} & i<j,\\[4pt]
\Gamma_i^\alpha a+_j\Gamma_i^\alpha b
    & \text{if} & i>j.
\end{array}
\right.
\end{equation}

For $i=j$, the two \textit{connection transport laws} are
\begin{equation*}
\begin{array}{l}
\Gamma_j^+(a+_j b)
 =
(\Gamma_j^+a+_j\varepsilon_{j+1}b)
 +_{j+1}(\varepsilon_j b+_j\Gamma_j^+b),\\[6pt]
\Gamma_j^-(a+_j b)
 =
(\Gamma_j^-a+_j\varepsilon_j a)
 +_{j+1}(\varepsilon_{j+1}a+_j\Gamma_j^-b).
\end{array}
\end{equation*}
Equivalently, with the matrix convention just specified,
\begin{equation*}
\Gamma_j^+(a+_j b)=
\left[
\begin{array}{lr}
\varepsilon_j b & \Gamma_j^+b\\
\Gamma_j^+a & \varepsilon_{j+1}b
\end{array}
\right]
\qquad
\xymatrix @-1pc @R=0.3cm @C=0.3cm {
j & *{} \ar[l] \ar[d]\\
*{} & j+1
}
\end{equation*}

\begin{equation*}
\Gamma_j^-(a+_j b)=
\left[
\begin{array}{lr}
\varepsilon_{j+1}a & \Gamma_j^-b\\
\Gamma_j^-a & \varepsilon_j a
\end{array}
\right]
\qquad
\xymatrix @-1pc @R=0.3cm @C=0.3cm {
j & *{} \ar[l] \ar[d]\\
*{} & j+1
}
\end{equation*}

\medskip

Supported by grants PAPIIT-UNAM IN114723 and IN119726

\medskip
 
\bibliographystyle{unsrt}

\bibliography{HLGFs1refs-2}

@misc{Orendain:2023tly,
    author = "Orendain, Juan and Zapata, Jose Antonio",
    title = "{Higher homotopy and lattice gauge fields}",
    eprint = "2311.02363",
    archivePrefix = "arXiv",
    primaryClass = "math.CT",
    howpublished = "arXiv:2311.02363 [math.CT]",
    url = "https://arxiv.org/abs/2311.02363",
    year = "2023"
}

@article{brown2007new,
  title={A new higher homotopy groupoid: the fundamental globular {$\omega$}-groupoid of a filtered space},
  author={Brown, Ronald},
  journal={Homology, Homotopy and Applications},
  volume={10},
  number={1},
  pages={327--343},
  year={2008},
  eprint={math/0702677},
  archivePrefix={arXiv},
  url={https://groupoids.org.uk/pdffiles/globularhha.pdf}
}

@article{Meneses:2017vqn,
    author = "Meneses, Claudio and Zapata, Jose A.",
    title = "{Homotopy classes of gauge fields and the lattice}",
    eprint = "1701.00775",
    archivePrefix = "arXiv",
    primaryClass = "math-ph",
    doi = "10.4310/ATMP.2019.v23.n8.a7",
    journal = "Adv. Theor. Math. Phys.",
    volume = "23",
    number = "8",
    pages = "2207--2254",
    year = "2019"
}

@article{Meneses:2019bok,
    author = "Meneses, Claudio and Zapata, Jose A.",
    title = "{Macroscopic observables from the comparison of local reference systems}",
    eprint = "1905.04797",
    archivePrefix = "arXiv",
    primaryClass = "gr-qc",
    doi = "10.1088/1361-6382/ab49a7",
    journal = "Class. Quant. Grav.",
    volume = "36",
    number = "23",
    pages = "235011",
    year = "2019"
}

@article{schreiber2007parallel,
  title={Parallel transport and functors},
  author={Schreiber, Urs and Waldorf, Konrad},
  journal={Journal of Homotopy and Related Structures},
  volume={4},
  number={1},
  pages={187--244},
  year={2009},
  eprint={0705.0452},
  archivePrefix={arXiv},
  url={https://arxiv.org/abs/0705.0452}
}

@book{kobayashi1996foundations,
  title={Foundations of Differential Geometry, Volume {I}},
  author={Kobayashi, Shoshichi and Nomizu, Katsumi},
  series={Wiley Classics Library},
  year={1996},
  publisher={John Wiley \& Sons},
  isbn={978-0-471-15733-5}
}

@article{Barrett:1991aj,
    author = "Barrett, John W.",
    title = "{Holonomy and path structures in general relativity and Yang-Mills theory}",
    doi = "10.1007/BF00671007",
    journal = "Int. J. Theor. Phys.",
    volume = "30",
    pages = "1171--1215",
    year = "1991"
}

@book{iglesias2013diffeology,
  title={Diffeology},
  author={Iglesias-Zemmour, Patrick},
  series={Mathematical Surveys and Monographs},
  volume={185},
  year={2013},
  publisher={American Mathematical Society},
  doi={10.1090/surv/185}
}

@book{BrownNAT,
  title={Nonabelian Algebraic Topology: Filtered Spaces, Crossed Complexes, Cubical Homotopy Groupoids},
  author={Brown, Ronald and Higgins, Philip J. and Sivera, Rafael},
  series={EMS Tracts in Mathematics},
  volume={15}, 
  year={2011},
  publisher={European Mathematical Society},
  doi={10.4171/083}
}

@article{Algebraofcubes,
 author="Brown, Ronald and Higgins, Philip J.",
 title="On the algebra of cubes",
 journal="Journal of Pure and Applied Algebra",
 volume="21",
 year="1981",
 pages="233--260",
 }

@article{whitehead1948operators,
  title={On operators in relative homotopy groups},
  author={Whitehead, J. H. C.},
  journal={Annals of Mathematics},
  volume={49},
  number={3},
  pages={610--640},
  year={1948},
  publisher={JSTOR}
}

@article{brown1979colimit,
  title={Colimit theorems for relative homotopy groups},
  author={Brown, Ronald and Higgins, Philip J.},
  journal={Journal of Pure and Applied Algebra},
  volume={22},
  pages={11--41},
  year={1981}
}

@article{Pfeiffer:2003je,
    author = "Pfeiffer, Hendryk",
    title = "{Higher gauge theory and a nonAbelian generalization of 2-form electrodynamics}",
    eprint = "hep-th/0304074",
    archivePrefix = "arXiv",
    reportNumber = "DAMTP-2003-27",
    doi = "10.1016/S0003-4916(03)00147-7",
    journal = "Annals Phys.",
    volume = "308",
    pages = "447--477",
    year = "2003"
}

@inproceedings{Orendain:2024wfh,
    author = "Orendain, Juan and Zapata, Jose A.",
    title = "{A better space of generalized connections}",
    booktitle = "{The Seventeenth Marcel Grossmann Meeting}",
    publisher = "World Scientific",
    pages = "2073--2086",
    year = "2026",
    doi = "10.1142/9789819839926_0167",
    eprint = "2407.17400",
    archivePrefix = "arXiv",
    primaryClass = "gr-qc",
    url = "https://doi.org/10.1142/9789819839926_0167",
    note = "arXiv:2407.17400 [gr-qc]"
}

@article{Chen:2024ddr,
    author = "Chen, Jing-Yuan",
    title = "{Instanton density operator in lattice QCD from higher category theory}",
    eprint = "2406.06673",
    archivePrefix = "arXiv",
    primaryClass = "hep-lat",
    doi = "10.21468/SciPostPhys.19.6.158",
    journal = "SciPost Phys.",
    volume = "19",
    number = "6",
    pages = "158",
    year = "2025"
}

@article{Zhang:2024sgm,
    author = "Zhang, Peng and Chen, Jing-Yuan",
    title = "{An explicit categorical construction of instanton density in lattice Yang-Mills theory}",
    eprint = "2411.07195",
    archivePrefix = "arXiv",
    primaryClass = "hep-lat",
    doi = "10.1007/JHEP06(2025)085",
    journal = "JHEP",
    volume = "06",
    pages = "085",
    year = "2025"
}

@article{BMW:2008jgk,
    author = "D{\"u}rr, S. and others",
    collaboration = "BMW",
    title = "{Ab-Initio Determination of Light Hadron Masses}",
    eprint = "0906.3599",
    archivePrefix = "arXiv",
    primaryClass = "hep-lat",
    doi = "10.1126/science.1163233",
    journal = "Science",
    volume = "322",
    pages = "1224--1227",
    year = "2008"
}


\end{document}